\documentclass[linenumbers]{elsarticle}
\usepackage{graphicx}
\usepackage{amsmath}
\usepackage{units}
\usepackage{amssymb}
\usepackage{subfigure}
\usepackage{fixltx2e} 
\usepackage{verbatim} 
\usepackage[numbers]{natbib} 

\usepackage{lineno}
\linenumbers 

\begin{document} 

\title{Design of a horizontal neutron reflectometer for the European Spallation Source}
\date{\today}

\author[hzb,bmbf]{D. Nekrassov\corref{cor1}} 
\ead{daniil.nekrassov@helmholtz-berlin.de}
\author[hzb,hd]{M. Trapp}
\ead{marcus.trapp@helmholtz-berlin.de}
\author[hzb,bmbf]{K. Lieutenant}
\ead{klaus.lieutenant@helmholtz-berlin.de}
\author[hzg,bmbf]{J.-F. Moulin}
\ead{jean-francois.moulin@hzg.de}
\author[ess]{M. Strobl}
\ead{markus.strobl@esss.se}
\author[hzb,bmbf]{R. Steitz}
\ead{roland.steitz@helmholtz-berlin.de}
\address[hzb]{Helmholtz-Zentrum Berlin, Hahn-Meitner-Platz 1, D-14109 Berlin, Germany}
\address[hd]{Ruprecht-Karls-Universit\"at, Im Neuenheimer Feld 253, 69120 Heidelberg, Germany}
\address[hzg]{Helmholtz-Zentrum Geesthacht Outstation at Forschungs-Neutronenquelle Heinz Maier-Leibnitz, 85747 Garching, Germany}
\address[ess]{European Spallation Source ESS AB, Box 176, 221 00 Lund, Sweden}
\address[bmbf]{German Work Package for the ESS Design Update}
\cortext[cor1]{Corresponding author}

\begin{abstract}

A design study of a horizontal neutron reflectometer adapted to the general baseline of the long pulse European Spallation Source (ESS) is presented. The instrument layout comprises solutions for the neutron guide, high-resolution pulse shaping and beam bending onto a sample surface being so far unique in the field of reflectometry. The length of this instrument is roughly 55 m, enabling $\delta \lambda / \lambda$ resolutions from 0.5\% to 10\%. The incident beam is focused in horizontal plane to boost measurements of sample sizes of $1 \times 1$ cm$^2$ and smaller with potential beam deflection in both downward and upward direction. The range of neutron wavelengths utilized by the instrument is 2 to 7.1 (12.2, ...) $\mathrm{\AA}$ , if every (second, ...) neutron source pulse is used. Angles of incidence can be set between 0$^{\circ}$ and 9$^{\circ}$ with a total accessible q-range from $4\times 10^{-3}$ $\mathrm{\AA^{-1}}$ up to 1 $\mathrm{\AA^{-1}}$. The instrument operates both in $\theta/\theta$ (free liquid surfaces) and $\theta/2\theta$ (solid/liquid, air/solid interfaces) geometry. The experimental setup will in particular enable direct studies on ultrathin films (d $\approx$ 10 $\mathrm{\AA}$) and buried monolayers to multilayered structures of up to 3000 $\mathrm{\AA}$ total thickness. The horizontal reflectometer will further foster investigations of hierarchical systems from nanometer to micrometer length scale (the latter by off-specular scattering), as well as their kinetics and dynamical properties, in particular under load (shear, pressure, external fields). Polarization and polarization analysis as well as the GISANS option are designed as potential modules to be implemented separately in the generic instrument layout. The instrument is highly flexible and offers a variety of different measurement modes. With respect to its mechanical components the instrument is exclusively based on current technology. Risks of failure for the chosen setup are minimum.

\end{abstract}

\maketitle

\section{Introduction and science case}

Soft matter and life science systems investigated by neutron reflectometry (NR) continuously increase in complexity -- both in structure as well as in the number and the specific roles of their components. The same is true for hard matter systems. This development presents a permanent challenge and demand for continuous improvement of the neutron facilities. On the other hand there is a constantly increasing request for NR measurements, in particular in soft matter and life sciences, which stems from the unique possibilities of neutrons in probing amphiphilic and self-organizing structures at air-liquid \cite{Bib:Bradbury, Bib:Bauer, Bib:Yang}, liquid-liquid \cite{Bib:Campana, Bib:Zarbakhsh} and solid-liquid (buried) interfaces \cite{Bib:Dante, Bib:Watkins, Bib:Cardenas, Bib:Hellsing, Bib:Tatur}. NR in combination with off-specular and grazing-incidence small angle neutron scattering (GISANS \cite{Bib:Muller-Buschbaum}, SESANS \cite{Bib:Bouwman}, SERGIS \cite{Bib:Dalgliesh, Bib:Pynn, Bib:Vorobiev}) provides deep insight in self-assembly- and aggregation processes \cite{Bib:Griffiths}, 1-3d ordered interfacial films \cite{Bib:Muller-Buschbaum, Bib:Muller-Buschbaum2, Bib:Xia, Bib:Parnell} and the interplay of length scales in hierarchically structured systems \cite{Bib:Ruderer}. \\

Current research at the forefront of science with NR covers investigations of protein organization in bio-membranes \cite{Bib:Brun, Bib:Holt}, structure, organization and functioning of living cells at interfaces \cite{Bib:Huth}, nano-engineering \cite{Bib:Yang2, Bib:Zhuk, Bib:Pichon}, new materials and interface properties for energy (storage) applications \cite{Bib:Kalisvaart, Bib:Jerliu}, magnetic fluctuations and domain propagation \cite{Bib:Brussing} as well as magnetic ordering and the understanding of relevant length scales and energies \cite{Bib:Hjorvarsson}.\\

Complementary (on-board) in-situ techniques in addition to NR are often needed to elucidate complex interfacial structures and processes that are difficult to assess otherwise. In particular X-ray reflectometry \cite{Bib:Yang, Bib:Dabkowska}, ellipsometry \cite{Bib:Wallet}, Brewster angle microscopy \cite{Bib:Hollinshead, Bib:Bauer2} and IR spectroscopy \cite{Bib:Bauer2, Bib:Eastman, Bib:Leitch, Bib:Strobl, Bib:Kreuzer} are essential complementary tools for achieving integrated insight into soft matter and life science systems. Soft matter sample sizes required by NR today are typically of the order of 1000 mm$^2$ or larger which hinders studies with components that are not available or cannot be produced in sufficient quantities. The latter holds in particular, but not exclusively, for biological systems where typically human proteins provide such bottle-neck. Thus, parametric studies for instance on biological, genetic or pharmaceutical activity are outside the scope of NR measurements today. Such investigations will become available with the high flux of the ESS source by utilizing much reduced beam cross sections (foot prints) and sample sizes. \\

Minimized beam dimensions (with not necessarily minimized sample size) will further allow for characterizations of rough and curved interfaces that are common in functional and natural materials like industrial coatings, ball bearings, bone or skin. In addition, inhomogeneous, patterned surfaces might be scanned for local features with sub-mm beams. Magnetic reference layers in combination with polarized neutron beams offer additional contrast and sensitivity for samples that do not withstand solution contrast changes \cite{Bib:Holt}. In combination with polarization analysis such experimental set-up will help discriminating the incoherent background in soft matter and life science systems by spin-filtering \cite{Bib:Ioffe}.  \\

The ESS source \cite{Bib:ESS} will enable kinetic studies of systems in non-equilibrium situations on time scales not available today. Tackling the millisecond range in the future however might still require concurrent sample sizes. Therefore, the reflectometer at ESS should foresee flexible and adaptable collimation optics. The latter is also important whenever interfaces are to be assessed from above and below the sample horizon as can be the case for buried interfaces and interfaces under load, e.g. shear forces in combined NR and rheology set-ups \cite{Bib:Wolff, Bib:Gutfreund}. \\

The length scale of interest in soft matter and life science systems of the next generation will cover 1-3000 $\mathrm{\AA}$, hence an instrument with variable resolution from $0.5\% \leq dq/q \leq 10\%$ is required. For the majority of purposes (not including wide angle and diffraction experiments) a q-range of 0.005 $\mathrm{\AA}^{-1}$ to 1 $\mathrm{\AA}^{-1}$ will suffice. \\

\section{Layout of the horizontal reflectometer}

The scientific motivation described in the last section combined with additional requirements from ESS advisory panels were taken into account when choosing the basic instrument parameters. As a result, the key strength of this instrument was decided to be the ability to use small samples and to access large momentum transfer values $q$ reaching up to $\approx$ 1 $\mathrm{\AA^{-1}}$ in both downward and upward direction for liquid samples. In order to match these requirements, the following basic parameters were selected:

\begin{itemize}

\item Moderator: The horizontal reflectometer extracts neutrons from the cold moderator providing the highest flux in the region of interest $\lambda \geq 2 \mathrm{\AA}$.

\item Instrument length $L_{\mathrm {tot}}$: The instrument length is determined by the loosest desired resolution of $\delta \lambda / \lambda = 10 \%$ and shortest design wavelength currently being $2 \mathrm{\AA}$. Based on the ESS pulse length of $\Delta t = 2.86$ ms, $L_{\mathrm{tot}}$ is calculated from: 
\begin{equation*}
\frac{\Delta t}{L_{\mathrm{tot}}/v(\lambda = 2 \mathrm{\AA})} = 10 \%
\end{equation*}
which gives $L_{\mathrm{tot}} = 56.6$ m. The current length of the instrument design is 54.9 m and thus very close to the analytical value.

\item Usable waveband $\Delta \lambda$: Using the instrument length $L_{\mathrm{tot}} = 54.9$ m, the pulse length $\Delta t = 2.86$ ms and the pulse frequency $f = 14$ Hz, the width of the waveband $\Delta \lambda$ before main frames start to overlap is: $\Delta \lambda = t_\mathrm{eff} \times v_0 / L_{\mathrm{tot}} = 5.1 \, \mathrm{\AA}$, where $t_\mathrm{eff} = 1/f = 71.43$ ms and $v_0 = 3.956 \mathrm{\, m \, \mathrm{\AA}\, / \, ms}$. Thus wavelengths from 2 $ \mathrm{\AA}$ to 7.1 $ \mathrm{\AA}$ are used if every neutron pulse is accepted. The width of the waveband can be increased to $\approx 12$ ($\approx 17$, ...) $ \mathrm{\AA}$, if every second (third, ...) pulse is utilized. In principle, the used waveband can be optimized for each measurement, by shifting it to lower or higher wavelengths (e.g, 1.5 $ \mathrm{\AA}$ - 6.6 $ \mathrm{\AA}$ or 4  $ \mathrm{\AA}$ to 9.1 $ \mathrm{\AA}$) to improve the intrinsic resolution or q-range, respectively.

\item Vertical guide shape: A horizontal reflectometer is to provide only a limited vertical divergence on the sample, which depends on the desired $\delta q / q$ resolution. Hence, a constant guide shape can be chosen in the vertical plane, which does not provide large divergence on the sample and is easy to build. At the same time, in order to achieve high $q$ values of up to 1 $\mathrm{\AA^{-1}}$ in the case of free liquid surfaces, the beam needs to be bent onto the sample surface by up to $ \theta_i \approx$ 9$^\circ$. Such a severe beam deflection is very challenging and requires multiple reflections from tilted or curved mirror surfaces, which is best managed with limited beam heights. Thus the height of the beam $h_0$ is restricted to 2 cm. The beam deflection part of the guide is described in the next section.

\item Horizontal guide shape: It is expected that sample sizes for reflectometry will tendencially decrease. To provide a sufficient neutron flux on samples of the order of 1 cm$^2$, the guide system should have focusing properties in the horizontal plane, hence the reflectometer is currently planned with an elliptic guide \cite{Bib:ESSGuides, Bib:EG1}, starting with entrance/exit width of 10 cm and a maximum width of 26 cm. The elliptic guide provides a focused beam to either perform efficient measurements with small samples or restrict the beam size to carry out position dependent studies. Still, the outgoing beam is large enough for also addressing bigger samples, e.g. for time-resolved measurements. The distance between the source and the guide entrance is 2 m being currently the minimum distance, at which optical components can be placed at the ESS. The first 4 m part of the guide is thus inside the shielding monolith of 6 m radius and is referred to below as the \textbf{extraction system} (guide part 1).

\begin{figure}
\begin{center}
	\subfigure[Top view of the horizontal reflectometer]{\includegraphics[width=0.99\textwidth]{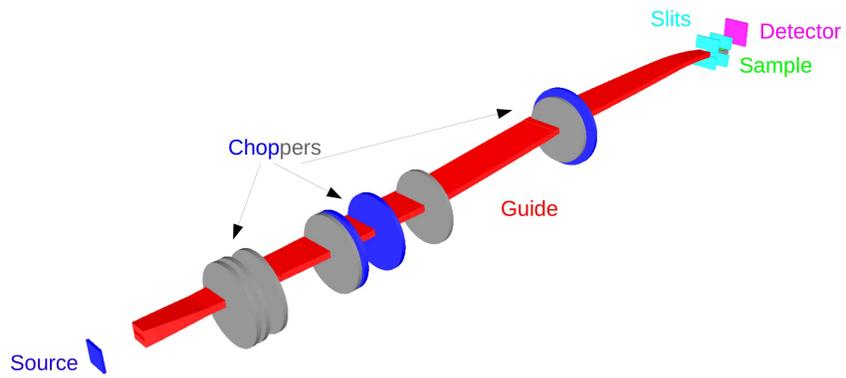}}
	\subfigure[Side view of the horizontal reflectometer]{\includegraphics[width=0.99\textwidth]{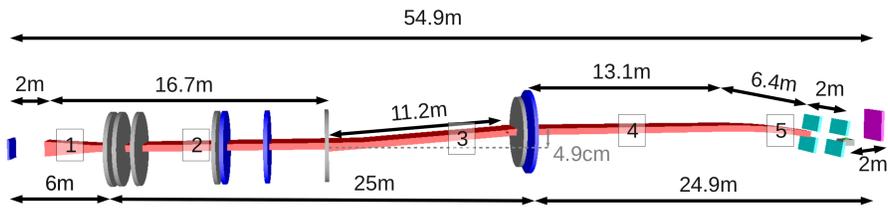}}
	\caption{A sketch of the overall layout of the reflectometer. For positions and dimensions of individual components see Tab. \ref{Tab:Reflectometer}. In the side view picture, different guide segments are denoted by numbers, see also Tab. \ref{Tab:Reflectometer}. The choppers used in the basic setup are colored in blue, while the ones used for the WFM high-resolution setup are grey. }
\label{Fig:LayoutWFM}
\end{center}
\end{figure}

\end{itemize}

In the following, the layout of the instrument based on the key parameters listed above is described in more detail (see also Tab. \ref{Tab:Reflectometer} for an overview). Individual beamline components are presented and their characteristics are justified based on results of neutronic MC simulations carried out with the VITESS software package \cite{Bib:Vitess}.

\subsection{Moderator and extraction system}

The reflectometer is set up for using wavelengths of 2 $\mathrm{\AA}$ and longer. Based on the current moderator description provided by the ESS and implemented in simulation packages like VITESS \cite{Bib:Vitess} and McStas \cite{Bib:Mcstas}, the cold moderator with 12 cm edge length, which provides a larger flux for wavelengths longer than 2.5 $\mathrm{\AA}$, was selected for the reflectometer. Neutron optic elements can be placed in the shielding monolith, which is currently 6 m in radius, beginning at 2 m behind the moderator. Thus beamline components like mirrors or guides can be placed in the monolith to increase the number of neutrons fed into the main guide system outside the first 6 m. While the horizontal shape is elliptic (see above), three options were tested for the vertical shape of the extraction system, that is a constant height of 2 cm, and a tapered guide with the entry height of 8.67 cm and 12 cm, respectively. For the latter, the guide entry height matches the size of the moderator. Simulations reveal that the best beam extraction is provided by the tapered feeding guide with 8.67 cm entry height, though it is noted that the performance of the constant guide is only slightly worse. The conclusion is that a large part of additional phase space accepted by the tapered guide is not transported by the subsequent beamline to the sample position.

\begin{figure}
\begin{center}
	\includegraphics[width=0.98\textwidth]{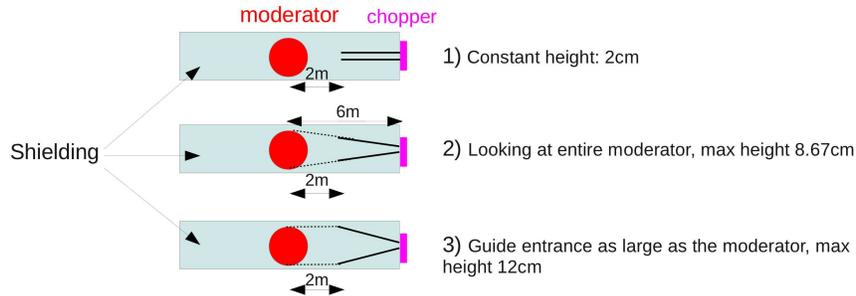}	
	\caption{Sketch of the three options studied for the vertical geometry of the extraction system within the main shielding monolith.}
\label{Fig:ExtSystemDrawing}
\end{center}
\end{figure}

\begin{figure}
\begin{center}
	\subfigure[Vertical divergence at 9 m]{\includegraphics[width=0.47\textwidth]{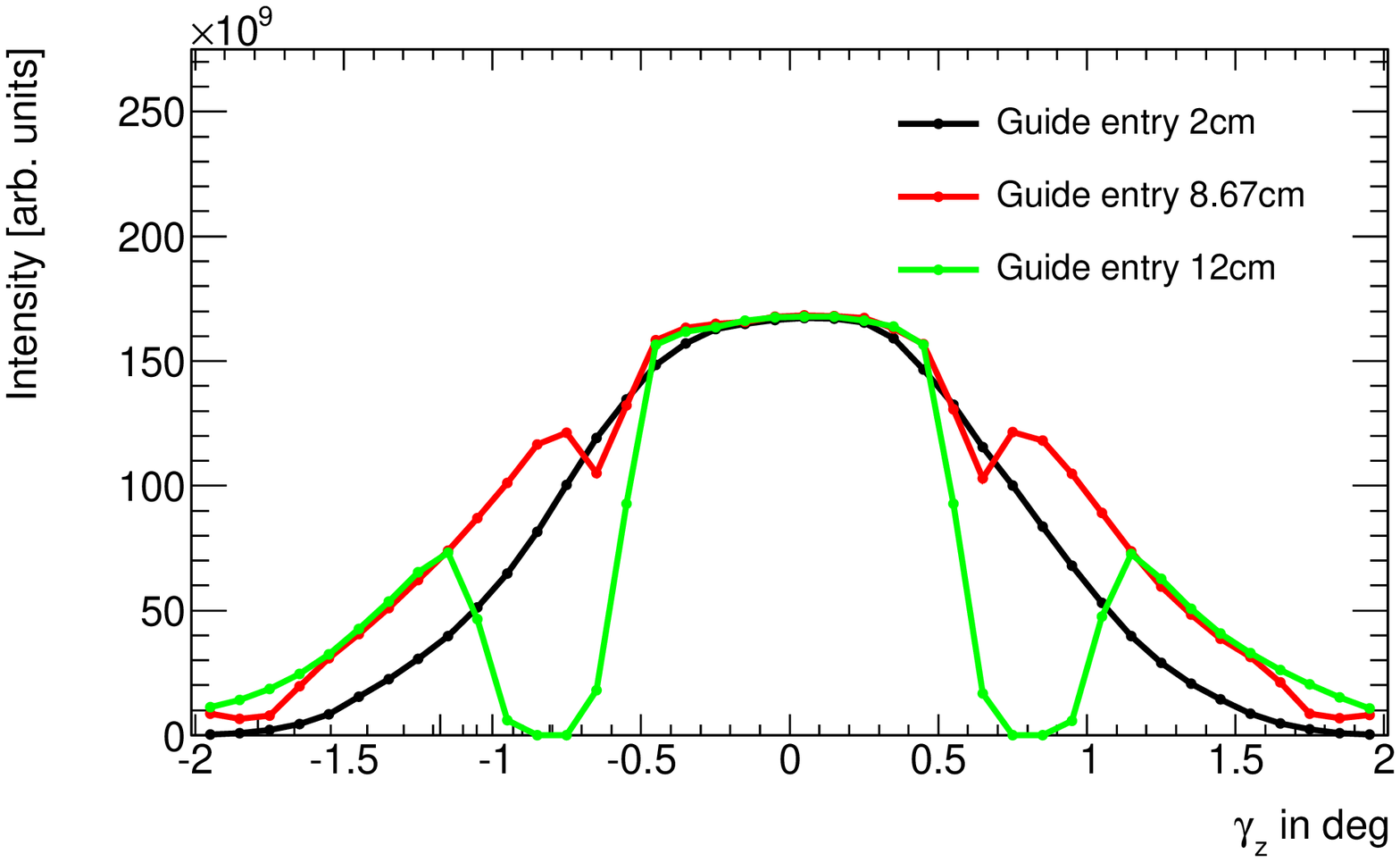}}
	\subfigure[Spectrum at the FPS position]{\includegraphics[width=0.47\textwidth]{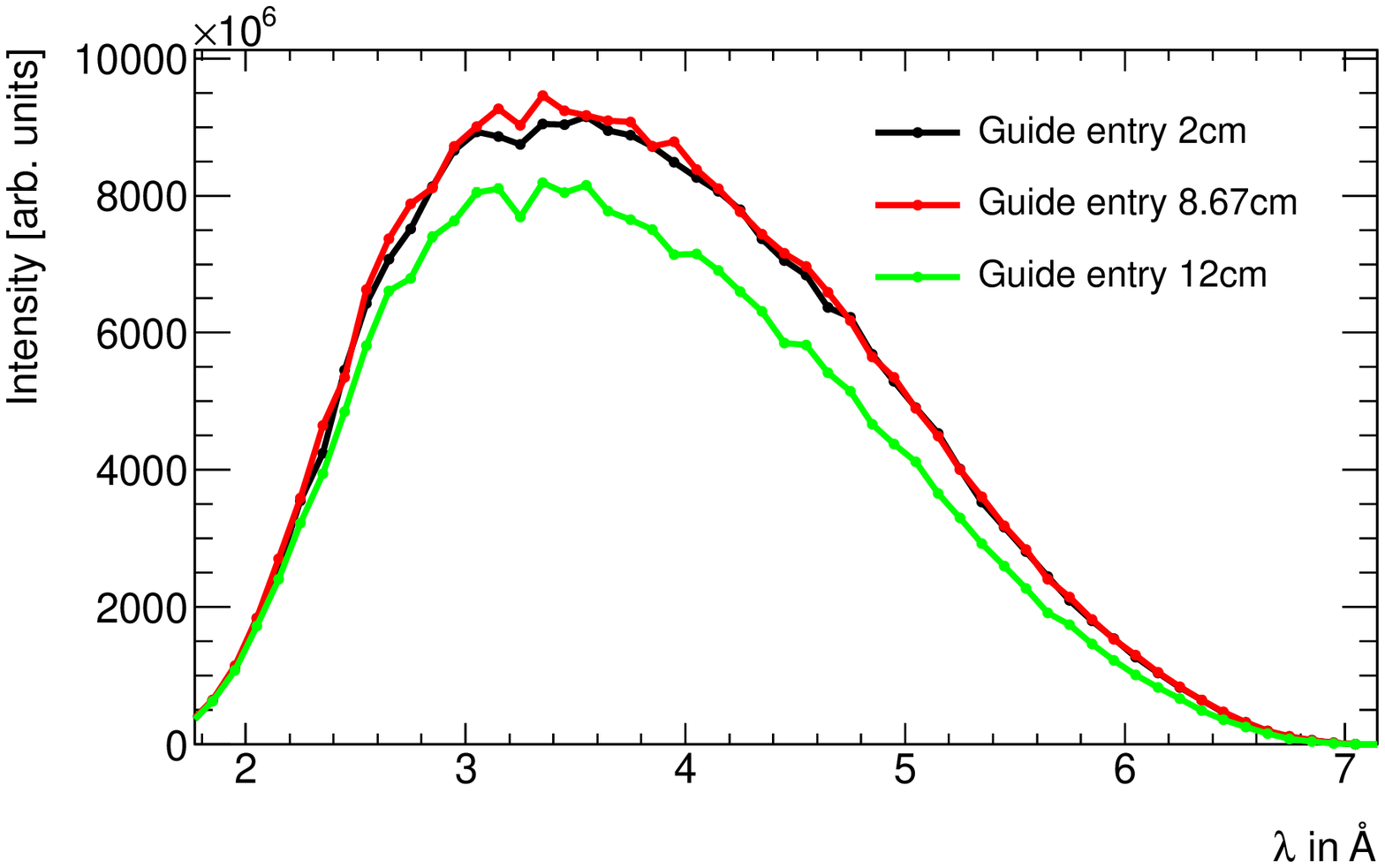}}
	\caption{Performance comparison of three different solutions for the vertical shape of the extraction system. a) The vertical divergence distribution was measured 3 m downstream of the extraction system. The studied solutions show differences in the divergence distribution for $|\gamma_z| > 0.5^{\circ}$. b) The intensity as a function of wavelength was measured before the footprint slit at 52.9 m downstream of the moderator for a bending angle of 2$^{\circ}$. The tapered guide option with 8.67 cm guide entry offers a slightly better performance than a constant height of 2 cm. The solution with the guide entry of 12 cm exhibits the worst performance of the three options studied.}
\label{Fig:ExtSystemComparison}
\end{center}
\end{figure}

\subsection{Direct line of sight}

A reflectometer aiming for high-q measurements needs to be free of prompt pulse background. Taking into account the unprecedented intensity of currently 5 MW and the energy of the primary proton beam of $ \approx 2.5$ GeV it is unclear whether a T0-chopper is able to remove all of the fast particle and gamma-ray background. This is why a T0-chopper is not foreseen and the background from the prompt pulse is avoided by the geometry of the guide system. The latter has to be designed such that in the field of view of the detector, there are no guide elements that lie in the direct flight path from the source. This is the definition of \textbf{avoiding the line of sight twice}. Since the horizontal shape of the guide is elliptic and a modification of this shape would interfere with its focusing properties, avoiding the line of sight must be carried out in the vertical plane.\\

\begin{figure}
\begin{center}
	\subfigure[Vertical divergence after leaving the line of sight]{\includegraphics[width=0.47\textwidth]{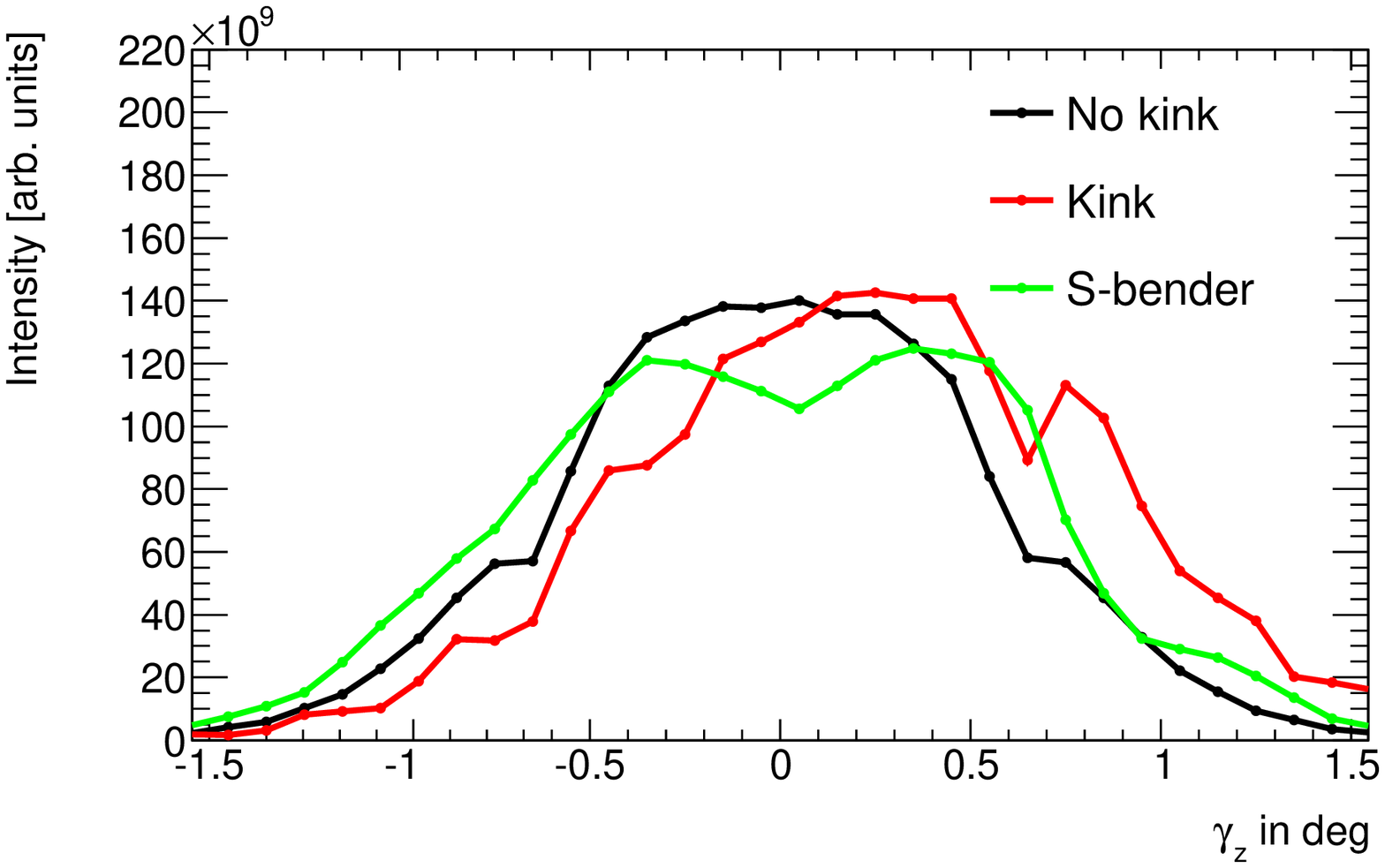}}
	\subfigure[Spectrum at the sample position]{\includegraphics[width=0.47\textwidth]{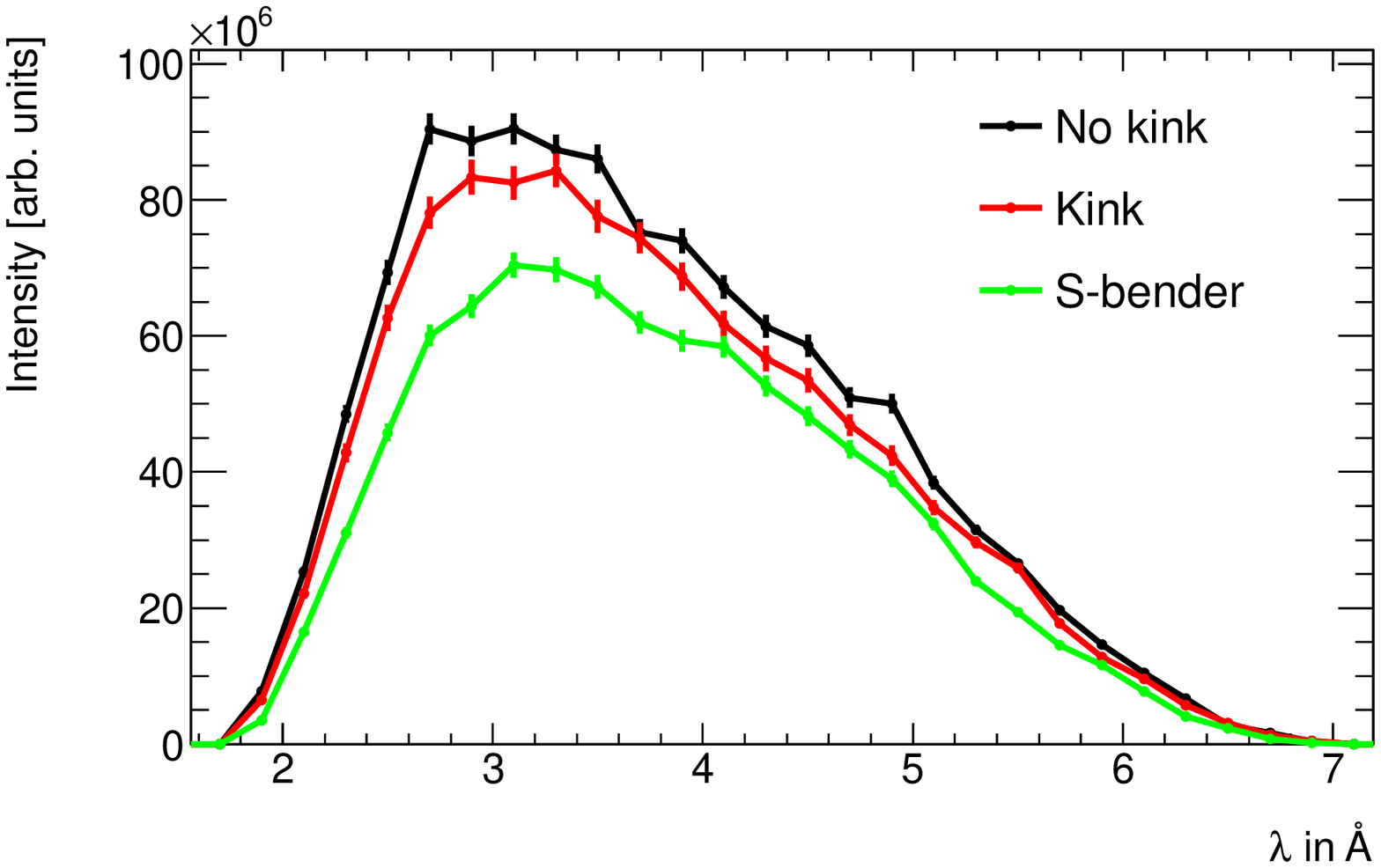}}
	\caption{(a) Comparison of the divergence distributions between instrument setups not avoiding the line of sight (no kink), or including an s-bender or z-kink mounted at 18.7 m downstream of the moderator. The z-kink option provides more flux (b) at the sample position due to a higher fraction of neutrons with low divergence after the z-kink, if compared with the S-bender option.}
\label{Fig:LoSBasicSetups}
\end{center}
\end{figure}

As a sophisticated chopper system will be placed within the first few meters after the shielding monolith, where also one chopper needs to be moved along the beamline, it was decided to leave the in-monolith section horizontally, i.e. not to have any inclination in the guide system there. On the other hand, the guide section 4, which is upstream from the bending section, should be horizontal, to enable the latter to equally access samples from above and below. Avoiding the line of sight requires a difference in height between the guide position at the monolith and at the bending section. Ways to achieve that are to place a double bender (\textbf{s-bender}) or a double kink (\textbf{z-kink}) in between. Here, a z-kink shows a clearly superior performance, see Fig. \ref{Fig:LoSBasicSetups}. The kink angle is fixed at $\alpha_{\mathrm K} = 0.25^{\circ}$, which leads to very moderate losses only if compared with a guide system without a kink. The shift of the maximum in the vertical divergence distribution towards 0.25$^\circ$ can be reduced by optimizing the vertical shape of the kink guide element. The latter shows the best performance if the shape is slightly double-elliptic. The total length of the kink guide section is 11.2 m and the shift in height of the guide position amounts to $\approx 5$ cm. \\

In accordance with the demand of avoiding the line of sight twice, the z-kink element should be placed such that fast primary particles are unable to illuminate the guide section after the z-kink, independent of whether their flight path would interfere with the detector area. A simple raytracing study was performed to determine an optimal position of the latter guide element by scanning the neutron parameter space at the source between $-6 \, \, \mathrm{cm} \leq z_\mathrm{S} \leq 6 $ cm and $-2^\circ \leq \gamma_z \leq 2^\circ$, where $z_\mathrm{S}$ is the vertical coordinate of the neutron trajectory at the source and $\gamma_z$ is its vertical divergence. The study revealed that if the z-kink is placed at a distance of 18.7 m away from the source, all fast particles would propagate at least 6 m outside the guide before entering the last guide section. Since propagation outside the guide can for simplicity be considered as propagation in shielding, it can safely be argued that due to the guide geometry, all fast background will be sufficiently absorbed and the prompt pulse will not contribute to the background at the detector position. \\

\subsection{Beam delivery on sample}

\begin{figure}
\begin{center}
	\subfigure[Horizontal beam profile at footprint slit]{\includegraphics[width=0.47\textwidth]{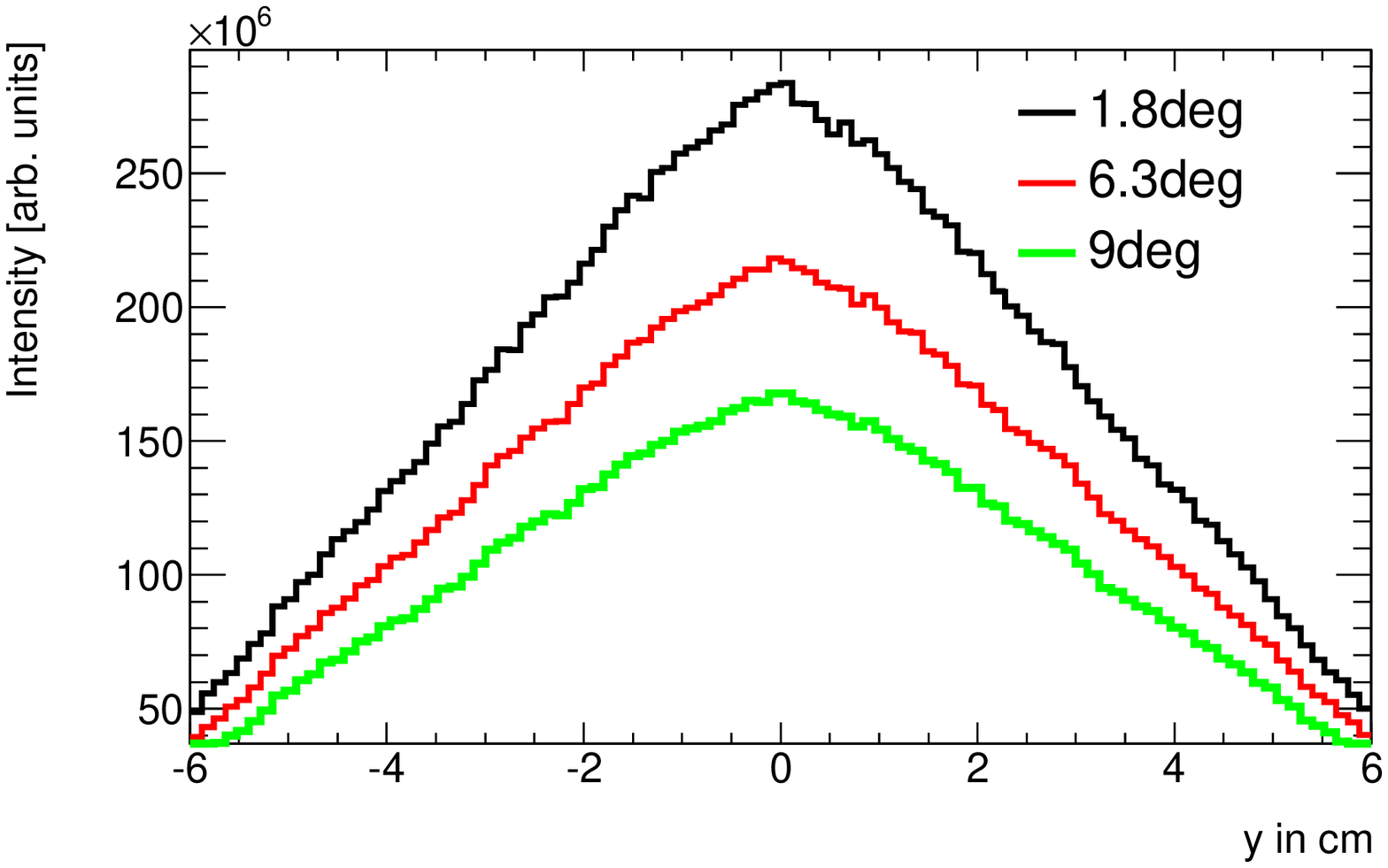}}
	\subfigure[Vertical beam profile at footprint slit]{\includegraphics[width=0.47\textwidth]{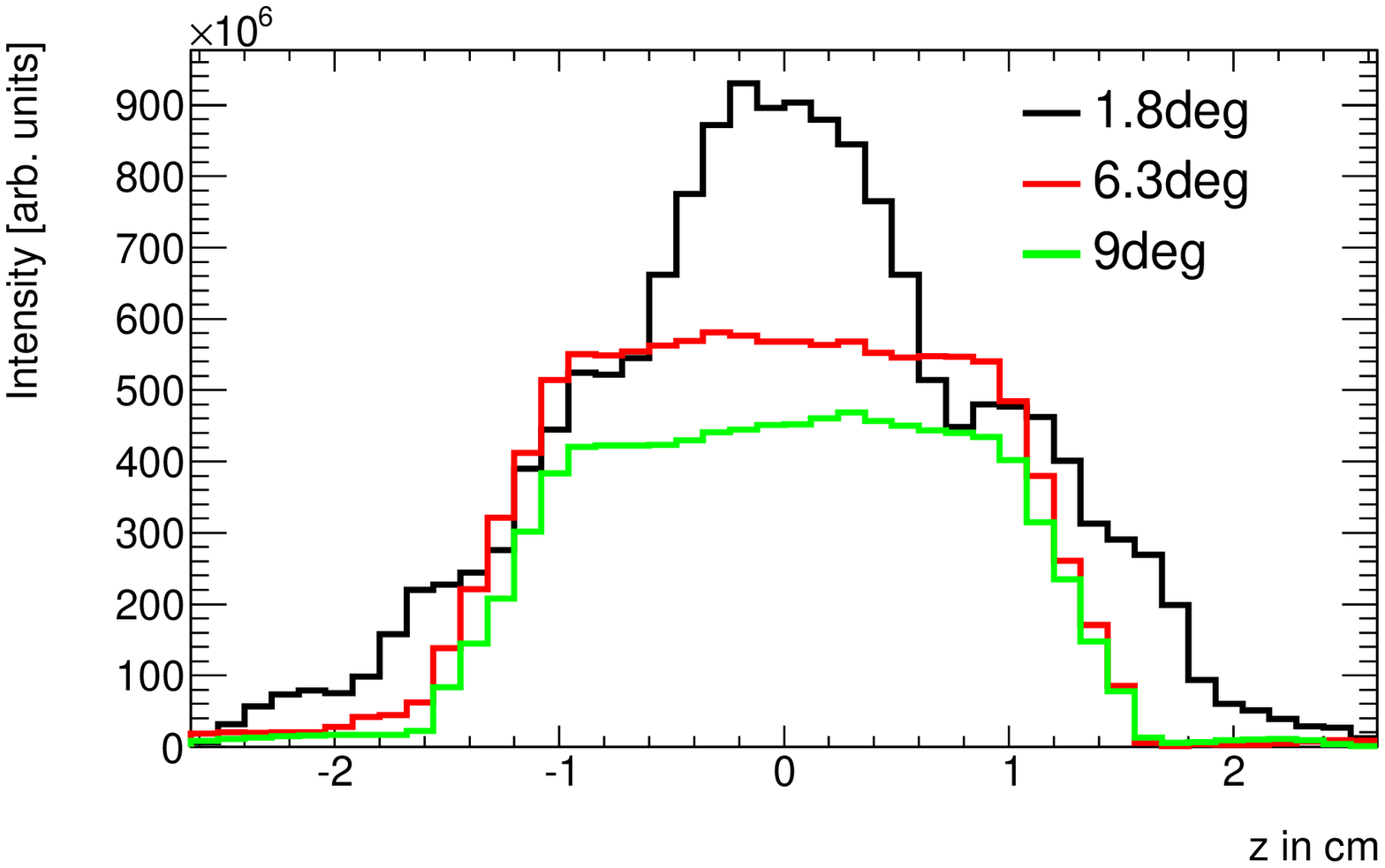}}
	\caption{Beam profile measured perpendicular to the beam direction before footprint slit in horizontal and vertical direction for three different bending angles.}
\label{Fig:BeamPosAtSample}
\end{center}
\end{figure}

\begin{figure}
\begin{center}
	\subfigure[Vertical beam divergence at footprint slit]{\includegraphics[width=0.47\textwidth]{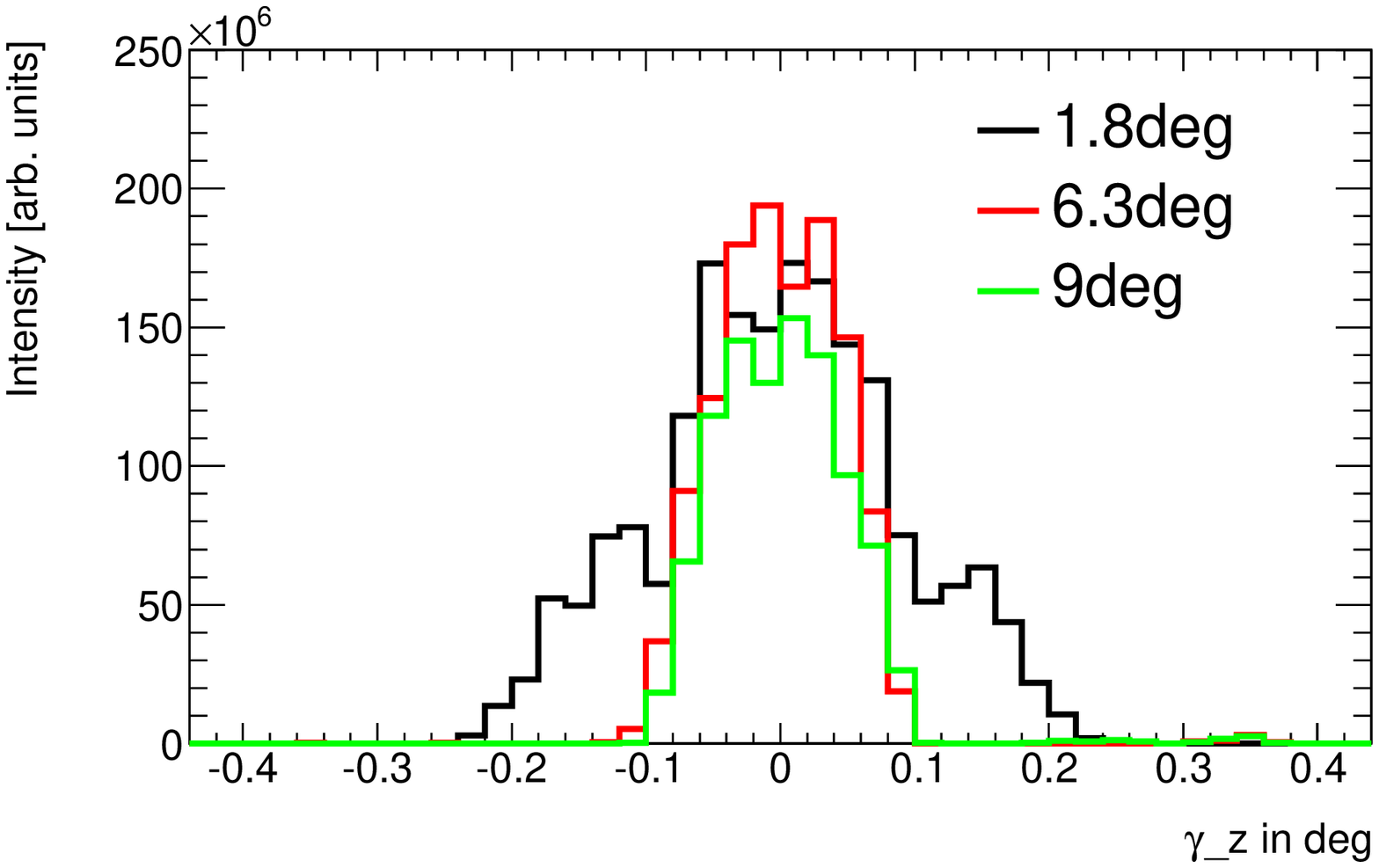}}
	\subfigure[Vertical beam divergence after footprint slit]{\includegraphics[width=0.47\textwidth]{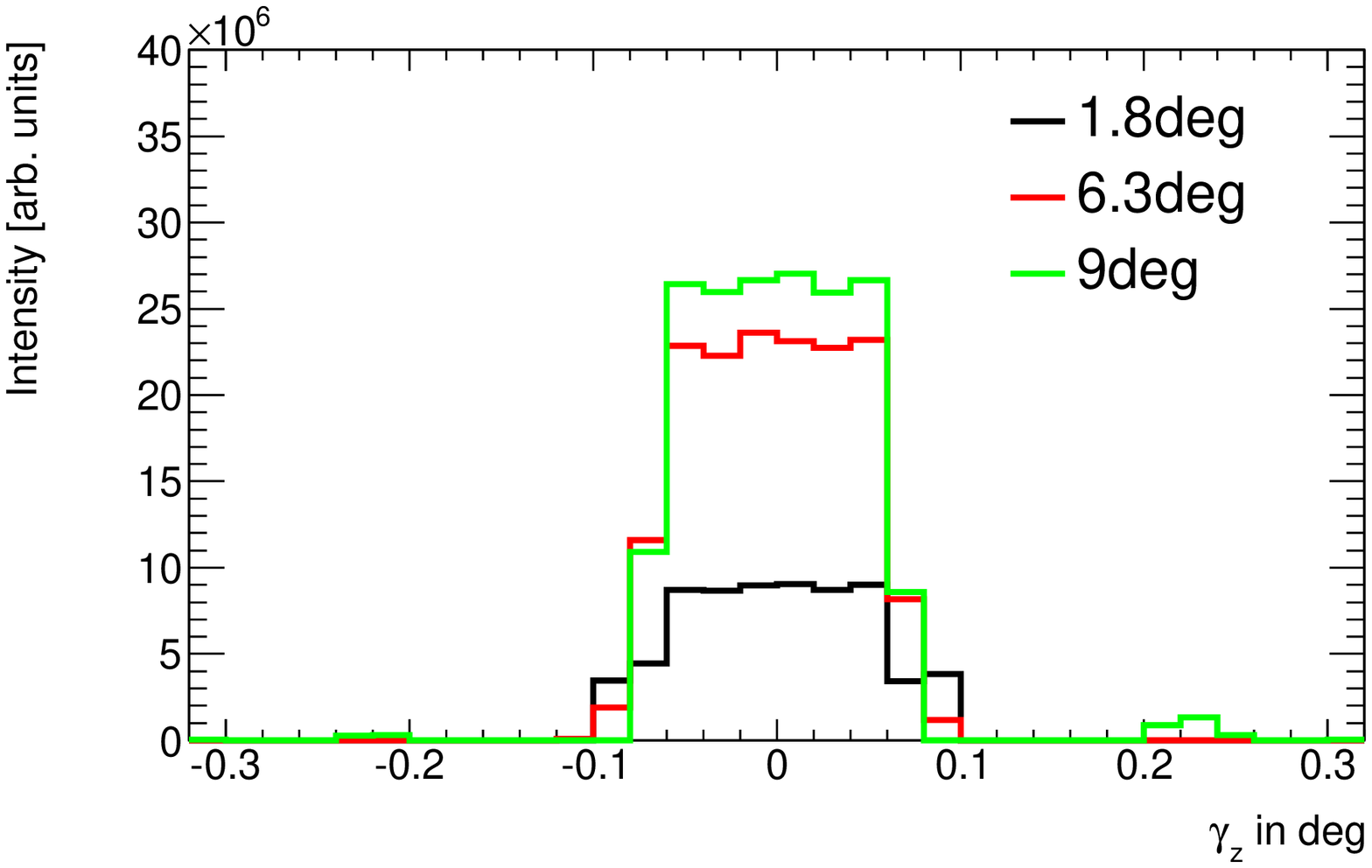}}
	\caption{Vertical divergence profile before footprint slit and at sample position for three different bending angles. The divergence profile before the footprint slit was measured perpendicular to the beam direction within $-1 \, \mathrm{cm} \leq y \leq 1 \, \mathrm{cm}$ and $-0.5 \, \mathrm{cm} \leq z \leq 0.5 \, \mathrm{cm}$. The first collimation slit was adjusted for a angular resolution $\delta \gamma_{z} / \gamma_z$ of 10\%. The size of the second (footprint) slit matches the footprint of a 1x1 cm$^2$ sample.}
\label{Fig:BeamDivAtSample}
\end{center}
\end{figure}

As stated above, in order to access a momentum transfer $q \approx 1 \, \mathrm{\AA^{-1}}$ on a free liquid surface the beam must have an incident angle on a horizontal sample of $\theta_\mathrm{tot} \approx 9^{\circ}$, taking into account the shortest utilized wavelength of 2 $\mathrm{\AA}$. Short wavelength neutrons require several bounces off the beam bending walls in order to be inclined by such large angle $\theta_\mathrm{tot}$, since there is a strong limitation on their maximum reflection angle, e.g. $\theta_\mathrm{max} = 0.1 m \lambda = 1^\circ$ for m=5 supermirror coatings. At the same time, the reflectometer is to provide sufficient flexibility concerning the variety of incident angles such that also regions of small $q \approx 0.005 \, \mathrm{\AA^{-1}}$ can be covered. Thus, the beam is tilted down- or upwards on a horizontal sample surface by means of 5 movable deflecting elements, of which the top and the bottom surfaces are covered with m = 5 coating. The length of each element is $\approx$ 1.28 m with the height of 2 cm. When tilted by $ \theta_i = 0.9^{\circ}$ each element bends the full beam of 2 cm height by 1.8$^\circ$. The horizontal shape of the beam bending elements is elliptic, since these elements are still part of the main guide ellipse (see Fig. 1). The maximum bending angle $\theta_\mathrm{tot} \approx 9^{\circ}$ guaranties for a maximum  $q \approx 1 \, \mathrm{\AA^{-1}}$. Bending guide elements can be replaced upon request by a GISANS guide module providing horizontal collimation. The beam characteristics after the bending section are shown in Figs. \ref{Fig:BeamPosAtSample} and \ref{Fig:BeamDivAtSample}. In principle, it is also possible to have shorter (1.15 m) bending elements that can be tilted by 1$^\circ$, hence reaching the maximum reflection angle for a 2 $\mathrm{\AA}$ neutron. On the other hand, MC simulations show that in this case the intensity at wavelengths around 2 $\mathrm{\AA}$ is heavily suppressed.\\

\begin{figure}
\begin{center}
	\includegraphics[width=0.98\textwidth]{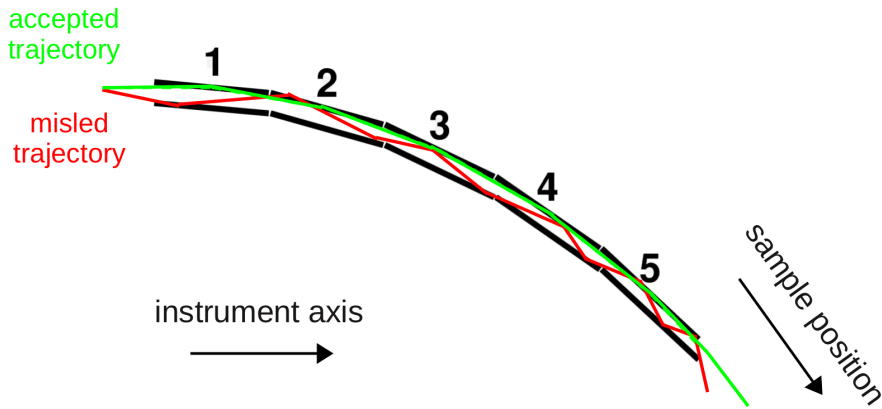}	
	\caption{Illustration of the bending principle used in the reflectometer. The five bending elements are arranged such that by reflections off each of the element's top (bottom) surface, the beam is deflected by the desired incident angle downwards (upwards) onto the sample. Neutrons with large divergence are likely to be deflected such that they either get absorbed or leave the bending section with even larger divergence, hence not propagating towards the sample.}
\label{Fig:BendingFig}
\end{center}
\end{figure}

The layout of the bending section is optimized for maximum flexibility and high-$q$ while keeping the complexity of the system as low as possible. The five bending elements can be arranged to bend the neutron beam down- or upwards by a number of possible angles. Their work principle is that neutrons bounce off each element once and propagate towards the sample afterwards, see Fig. \ref{Fig:BendingFig}. The drawback is that neutrons with a non-zero divergence have a certain chance of hitting the bottom (top) wall of one of the bending elements, if the beam is bent downwards (upwards). The resulting mis-orientation can be further increased during the propagation of the beam until the neutron is absorbed or leaves the bending section with a high divergence, i.e. does not propagate towards the sample. This is why the vertical divergence provided by the collimation system does not exceed 0.2$^\circ$ FWHM (0.3$^\circ$ as maximum) for higher bending angles, even though the maximum divergence can in principle achieve maximum $\approx 0.6^\circ$, taking into account the height of the guide of 2 cm and the collimation length of 2 m. It was investigated whether the divergence can be increased by involving more bending elements that would approximate a curved guide. This, however, would increase the mechanical complexity of the system, while MC simulations showed that there was no significant flux gain at the sample position. \\

\subsection{Collimation}

The collimation of the neutron beam exiting the bending section of the guide is done by means of two diaphragms (slits) located at the guide exit and 2 m downstream from the guide, respectively. The first slit determines the total vertical divergence on the sample, while the second, so-called footprint slit, is used to reduce the beam size such that it matches the sample dimensions or the requirements for the beam size, e.g., for position sensitive measurements. The adjusted divergence depends on the desired vertical angular resolution on the sample. For simplicity one usually uses matching resolution for divergence and wavelength, thus the opening (height) of the first slit depends on the pulse shaping regime used. The free propagation length of 2 m further allows to install, e.g., a polarization device between the collimation slits.\\

\subsection{Pulse shaping}

\begin{figure}
\begin{center}
	\includegraphics[width=0.98\textwidth]{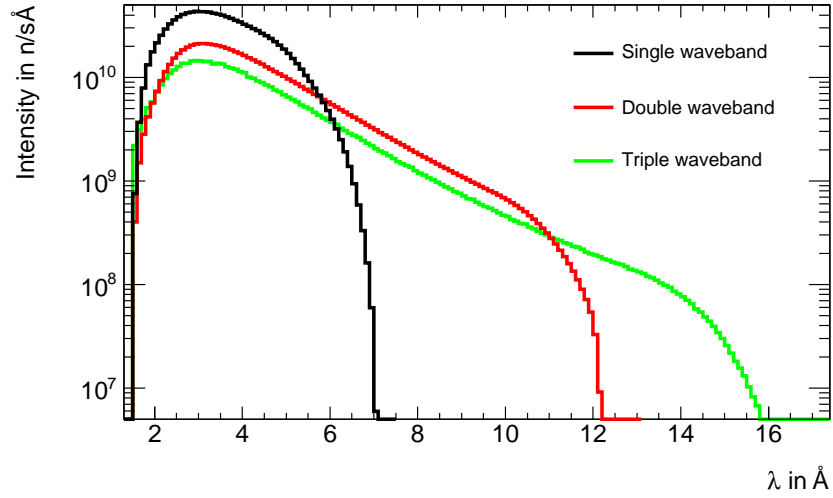}	
	\caption{The neutron spectrum as provided by the basic chopper layout of the reflectometer measured after the third chopper. By adjusting the rotation speed of the choppers the reflectometer can be provided with a single (double, ...) waveband anding from 2 $\mathrm{\AA}$ to 7.1 ($\approx 12$ \AA, ... ). Since only every second (third) pulse is used to extract the double (triple) waveband, for neutron wavelengths contained in the single waveband the flux is consequently reduced.}
\label{Fig:BasicSpectra}
\end{center}
\end{figure}

The reflectometer operates with two chopper setups for basic and high-resolution, respectively. All choppers are located at the side of the guide to make use of the small beam height and thus reduce the opening/closing time. The basic setup requires three choppers, which define the main frame and prevent neutrons with other wavelengths from propagating to the detector. In the basic setup, the wavelength resolution is defined by the total pulse length $t_0 = 2.86$ ms. Hence the first chopper does not need to be close to the source, as the pulse length will not be reduced. Furthermore, the first few meters after the shielding monolith are occupied by pulse shaping choppers belonging to the high-resolution setup (see below). Thus, the first chopper of the basic setup is placed at 12.5 m, followed by a second chopper at 17.5 m and a third chopper at 31 m. While the first two choppers are designed to suppress background from slow neutrons that otherwise could pollute the subsequent frame(s), the third chopper defines the accepted waveband of $2 - 7.1 \, \mathrm{\AA}$, if every source pulse is used. The waveband can be extended, if every second (third) pulse is used, see Fig. \ref{Fig:BasicSpectra}. This can be achieved by reducing the chopper rotation frequency to $1/2$ ($1/3$, ...) of the original value. \\

\begin{figure}
\begin{center}
	\includegraphics[width=0.98\textwidth]{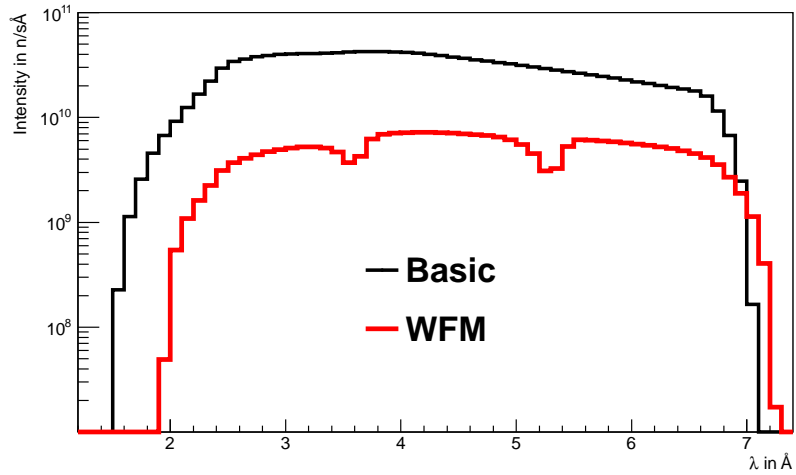}	
	\caption{Comparison between the neutron fluxes in front of Slit 1 (see Tab. \ref{Tab:Reflectometer} for position) using the basic and high-resolution setup, adjusted for a 1\% $\delta \lambda / \lambda$ resolution, for a reflection angle of 1.5$^\circ$. The intensity drop in the spectrum of the high-resolution system arises from strict prevention of subframe overlap at the detector position.}
\label{Fig:FluxBasicWFM}
\end{center}
\end{figure}

A dedicated wavelength frame multiplication system (WFM) was developed for the high-resolution work regime. Its purpose is to provide a high, constant wavelength resolution between $0.5 \%$ and $2.2 \%$ for the utilized waveband, while removing contaminant neutrons. This layout is very close to the one described in \cite{Bib:WFM} comprising six choppers in total. Solely the rotation speed of the first three choppers is increased to 112 Hz to further reduce intensity losses, see Fig. \ref{Fig:FluxBasicWFM} and Tab. \ref{Tab:Reflectometer}. It is noted that the intensity loss in subframe overlap regions is intrinsic to the way the WFM system is designed, as the subframes are kept separated in time very strictly. This is necessary for highly structured reflectivity spectra that are recorded with the high-resolution setup. In general, the ansatz described in \cite{Bib:WFM_Markus}, which is very similar to the one used here, could lead to less intensity losses in the subframe overlap regions, but at the expense of a somewhat more challenging subframe separation. \\

\subsection{Sample environment}

In case of free liquid surfaces or situations where the interface of study needs to be fixed in the horizontal plane, samples are put on a movable sample table equipped with motors for necessary movements in horizontal and in vertical direction. The positional displacement of the sample from the $z_0$ position of the main guide (upstream the bending section) is 84 cm at maximum for the highest possible deflection angle of 9$^{\circ}$. This maximum vertical movement is only 1.7 times the vertical displacement of the sample stage realized at FIGARO at present and thus, will not pose any difficulty in handling on the instrument. The sample table will allow for accessing the sample by the neutron beam from above and below. In the case of air-solid and solid-liquid interfaces samples will be tilted against the fixed incident beam in the horizontal at fixed zero height thus operating the instrument in the more simple $\theta/2\theta$ mode. In addition, there should exist enough space for sample environment controlling temperature, pressure, mechanical load etc. For the very reason the distance between the sample position and the footprint slit is set to 40 cm.

\subsection{Detector}

It is envisaged that the reflectometer comprises a position sensitive area detector of 50 x 50 cm$^2$, located 2 m away from the sample position, with a pixel size fine enough to match the angular resolution. The smallest reflection angles are expected to be 0.3$^\circ$ - 0.4$^\circ$. Taking into account the loosest wavelength resolution of 10\% the detector resolution should be $\approx 0.03^\circ$. Using $\Delta h/2 = \mathrm{tan}(\Delta \gamma_0^\circ)\times 2 \, \mathrm{m} \approx 1 $ mm, where $\Delta h$ is the pixel size and $\Delta \gamma_0 = 0.03^\circ$ the desired resolution expressed in $\sigma$ under the presumption that the divergence distribution is Gaussian, it follows that a detector with pixels having 2 mm egde length fulfills the instrument requirements. Furthermore, it will be advantegeous that the detection process takes place in a single conversion layer, since in this case the detector can be tilted by an angle $\alpha_\mathrm{det}$ w.r.t the neutron flight path after the sample to improve the angular resolution in vertical direction to $\Delta \gamma (\alpha_\mathrm{det}) = \Delta \gamma_0 \mathrm{cos} (\alpha_\mathrm{det})$. Such setup could even match the high wavelength resolution achieved with the WFM chopper setup.\\

\begin{table}[htbp]
 \begin{small}
  \begin{tabular}{|l|c|p{6.5cm}|}
    \hline
    \textbf{ Component }& \textbf{ Position [m] } & \textbf{Characteristics, parameters}   \\
    \hline	
    \hline	
	\multicolumn{3}{|c|} {Moderators} \\
     \hline
     Cold moderator & 0 & Liquid para-hydrogen, peak intensity at $\approx2.5$ \AA \\
    \hline
    \hline
     \multicolumn{3}{|c|} {Basic guide characteristics} \\
     \hline
     Guide & 2 & Length: 48.5 m, \newline Horizontal shape: elliptic with W$_i$ = W$_f$ = 10 cm, \newline W$_\mathrm{max}$: 26 cm, m=3 coating, \newline 
	moderator and footprint slit at focal points \newline
		 Vertical shape: variable, m=5 coating \\
    \hline                 	
    Guide part 1 (Extraction system) & 2 & L = 4 m, W$_i$ = 10 cm, W$_f$ = 16.6 cm, \newline
                                  H$_i$ = 8.67 cm, H$_f$ = 2 cm, linear shape\\
    \hline
    Guide part 2 & 6 & L = 12.7 m, W$_i$ = 16 cm, W$_f$ = 25.0 cm, \newline
                                         H$_i$ = H$_f$ = 2cm, vertical shape constant\\
    \hline
    Guide part 3 (z-kink) & 18.7 & L = 11.2 m,  W$_i$ = 25.0 cm, W$_f$ = 25.9 cm,\newline
	                                 H$_i$ = H$_f$ = 2cm, vertical shape double elliptic, \newline 
                                         H$_\mathrm{max}$ = 2.6 cm, H$_{1/2 \mathrm{L}}$ = 2 cm\\
    \hline
    Guide part 4 & 29.9 & L = 14.2 m, W$_i$ = 25.9 cm, W$_f$ = 19.1 cm,\newline
                                         H$_i$ = H$_f$ = 2cm, vertical shape constant\\
    \hline 
    Guide part 5 (bending section) & 44.1 & L = 6.4 m, composed of 5 tiltable sections of L$_s$ = 1.28 m\newline
		                  W$_i$ = 25.0 cm, W$_f$ = 25.9 cm\newline							
                                   H$_i$ = H$_f$ = 2cm, vertical shape constant	\\ 
    \hline
    \hline
     \multicolumn{3}{|c|} {Choppers} \\
     \multicolumn{3}{|c|} {Basic setup} \\
    \hline
     Basic chopper 1& 12.5 m & $\omega$ = 14 Hz, R = 40 cm, W$_i,f$ = 22 cm  \\
    \hline
     Basic chopper 2& 17.5 m & $\omega$ = 14 Hz, R = 40 cm, W$_i,f$ = 25 cm   \\
    \hline
     Basic chopper 3& 31 m & $\omega$ = 14 Hz, R = 40 cm, W$_i,f$ = 26 cm  \\
     \hline
     \multicolumn{3}{|c|} {WFM setup} \\
     \hline
     Pulse shaping chopper 1 & 6 & $\omega$ = 112 Hz, R = 35 cm, W$_i,f$ = 17 cm \\	
     \hline
     Pulse shaping chopper 2 & 6.25 m -- 7.08 m &$\omega$ = 112 Hz, R = 35 cm, W$_i,f$ = 17 -- 18 cm \\	
    \hline	
     Frame overlap chopper 1 & 7.5  &$\omega$ = 112 Hz, R = 35 cm, W$_i,f$ = 18 cm \\
     \hline
     Frame overlap chopper 2 & 12 & $\omega$ = 56 Hz, R = 40 cm, W$_i,f$ = 22 cm \\
     \hline
     Frame overlap chopper 3 & 19 & $\omega$ = 28 Hz, R = 40 cm, W$_i,f$ = 25 cm \\
     \hline
     Frame overlap chopper 4 & 30.4 & $\omega$ = 14 Hz, R = 40 cm,  W$_i,f$ = 26 cm \\
     \hline 	
     \hline
     \multicolumn{3}{|c|} {Collimation system} \\
     \hline
      Slit 1&  50.5 & W = 10 cm, height depends on required divergence \\
     \hline
      Slit 2&  52.5 & Width and height depend on sample size and incident angle \\	
     \hline
     \hline
      Sample table & 52.9 & Space for sample environment: 40 cm \\
      \hline
      Detector    & 54.9 & Position-sensitive pixel detector \newline
                           W$\times$H = 50 $\times$ 50 cm$^2$, pixel size 2 mm \\
      \hline
  \end{tabular}
  \end{small}
  \caption{Instrument components and their parameters used in the design of the reflectometer. Used symbols: Guide entry/exit width W$_i$/W$_f$ (also at chopper positions); Guide entry/exit height H$_i$/H$_f$; Maximum guide width/height W$_\mathrm{max}$/H$_\mathrm{max}$; Component length L; Angular chopper speed $\omega$; Chopper radius R; Detector width/height W/H. See text for further details.}\label{Tab:Reflectometer}
\end{table}

\subsection{SERGIS add-on}

\begin{figure}
\begin{center}
	\includegraphics[width=0.98\textwidth]{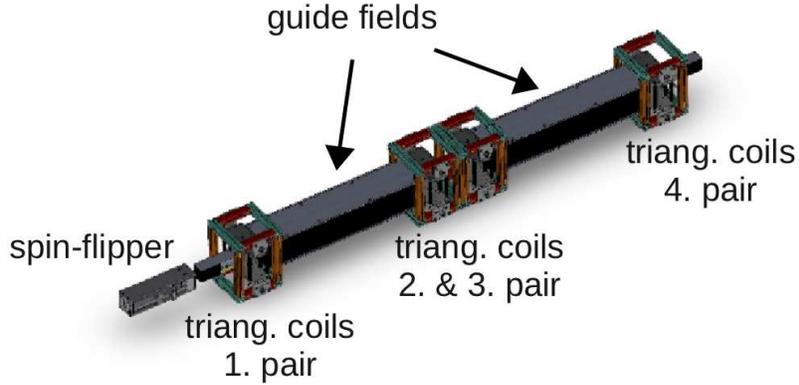}	
	\caption{Schematic respesentation of the SERGIS add-on. Shown are polariser, spin-flipper, triangular coils and guide fields.}
\label{Fig:SERGIS}
\end{center}
\end{figure}

As mentioned in the Introduction, interfacial systems continuously increase in complexity. While in the last decade main focus of neutron scattering investigations of interfaces was on 1-dimensional systems, this will change in near future with the development of hierarchical 2D- to 3D-ordered systems on nanoscale. Typical examples are porous interfacial films for use in catalysis, as scaffolds or templates for nanomaterial synthesis \cite{Bib:Yang2, Bib:Zhuk, Bib:Pichon}, as selective cell culture substrates \cite{Bib:Huth}, separation media, energy storage materials \cite{Bib:Kalisvaart, Bib:Jerliu}, spin valves or magnetic random access memory devices \cite{Bib:Brussing}. This development requires continuous improvement of adequate instrumentation for analysis. Here, SERGIS (Spin-Echo Resolved Grazing Incidence Scattering) \cite{Bib:Felcher, Bib:Major} as add-on offers a promissing route for the investigations of 2-3D structures on nanoscale. It utilizes the polarized beam of the instrument, can be mounted without alterations of instrument length into the reflectometry set-up and decouples the intensity of the incident beam from the resolution of the experiment. Although SERGIS is still not matured, in particular with respect to data interpretation and analysis \cite{Bib:Ashkar_2010,Bib:Ashkar_2011}, there is a great potential of this technique. In particular when used as rather simple \textbf{add-on}, and limited to nanoscale, functional systems, both in soft matter and life sciences \cite{Bib:Vorobiev} as well as in materials science \cite{Bib:Pynn}, also under non-equilibrium conditions might be explored with high spatial and temporal resolution.\\
The design of the SERGIS add-on is developed to fit the conditions imposed by the layout of the liquids reflectometer. For the setup 2 m in front and behind the sample position are available. Figure \ref{Fig:SERGIS} shows the technical design of the complete setup (polariser and analyser not shown). The installation at the incoming (upstream) side accommodates
\begin{enumerate}
	\item polariser (S-bender or polarized $^3$He gas cell), which is able to polarise a polychromatic beam with high efficiency.
	\item adiabatic gradient field radio frequency spin flipper for calibration and measurement possibility with both polarisations (spin up and spin down)
	\item	gradient-field coupling device (Forte coil) between the low field region (spin-echo precession devices, i.e. magnetic air-core coils with relatively weak fields) and the strong field region of the analyser
	\item first pair of triangular coils (A)
	\item guide field (GF1)
	\item second pair of triangular coils (B)
\end{enumerate}

The setup behind the sample (downstream) consists of:
\begin{enumerate}
	\item third pair of triangular coils (C)
	\item guide field (GF2)
	\item fourth pair of triangular coils (D)
	\item	gradient-field coupling device (Forte coil, (FC)) between the strong-field region (polarizer, flipper with strong permanent magnets) and the low-field region (spin-echo precession devices, i.e. magnetic air-core coils with relatively weak fields)
	\item analyser (multi-cavity) (Ana)
	\item position sensitive 2D detector (PSD)
\end{enumerate}

The guide field (GF1,2) is 1m in length in order to leave enough room on the incoming side to accommodate all spin manipulation devices. The angle $\chi$ of the triangular coils with respect to the incoming neutron beam can be changed between 30$^{\circ}$ and 60$^{\circ}$ depending on the orientation of the coils (see Figure \ref{Fig:SERGIS}). Magnetic fields used lie in the range from 1.5 to 15 mT. Above parameters in combination with a wavelength band of 2 - 7.1$\mathrm{\AA}$ (2 - 12$^{\circ}$, every second pulse) result in a spin echo length of 5-662 nm (30$^{\circ}$) and 9-1150 nm (5-1850 nm, 30$^{\circ}$ and 90-3200 nm, 60$^{\circ}$). 

\subsection{GISANS add-on}

\begin{figure}
\begin{center}
        \includegraphics[width=0.8\textwidth,keepaspectratio]{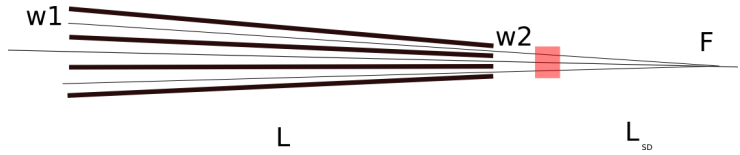}        
\caption{Schematics of the radial collimator with L = collimation length, LSD = sample-to-detector distance, F = focal point, w1, w2 = width of collimator channels at entrance, exit of the collimator.}
\label{Fig:GISANSLayout}
 \end{center}
\end{figure}

\begin{figure}
\begin{center}
        \includegraphics[width=0.8\textwidth,keepaspectratio]{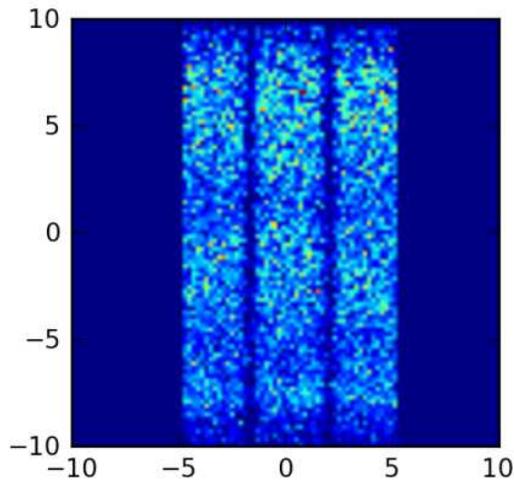}        
\caption{Illustration of expected 2D intensity distribution at the collimator exit.}
\label{Fig:GISANS2D}
 \end{center}
\end{figure}

\begin{figure}
\begin{center}
	\subfigure[]{\includegraphics[width=0.31\textwidth]{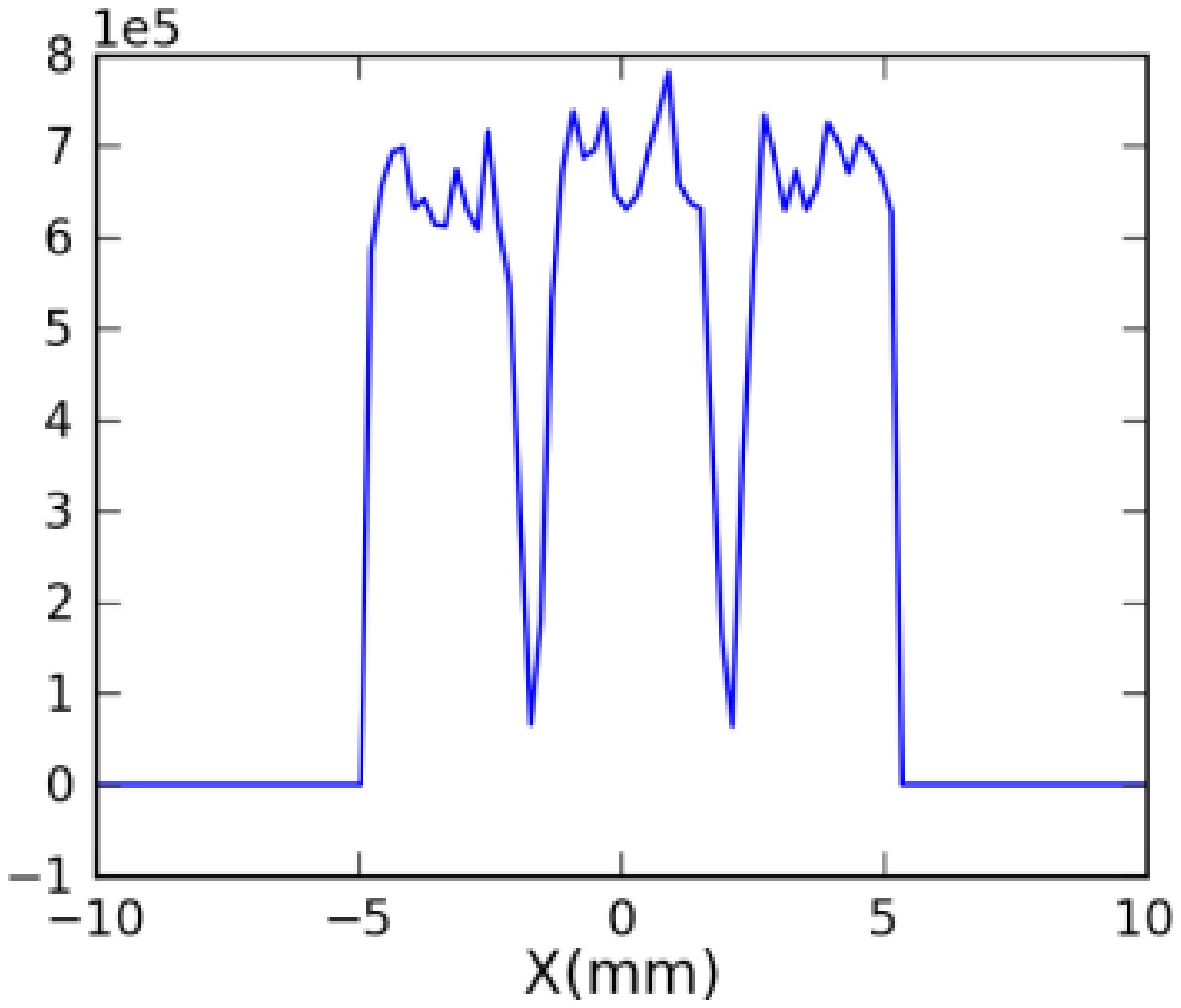}}
	\subfigure[]{\includegraphics[width=0.31\textwidth]{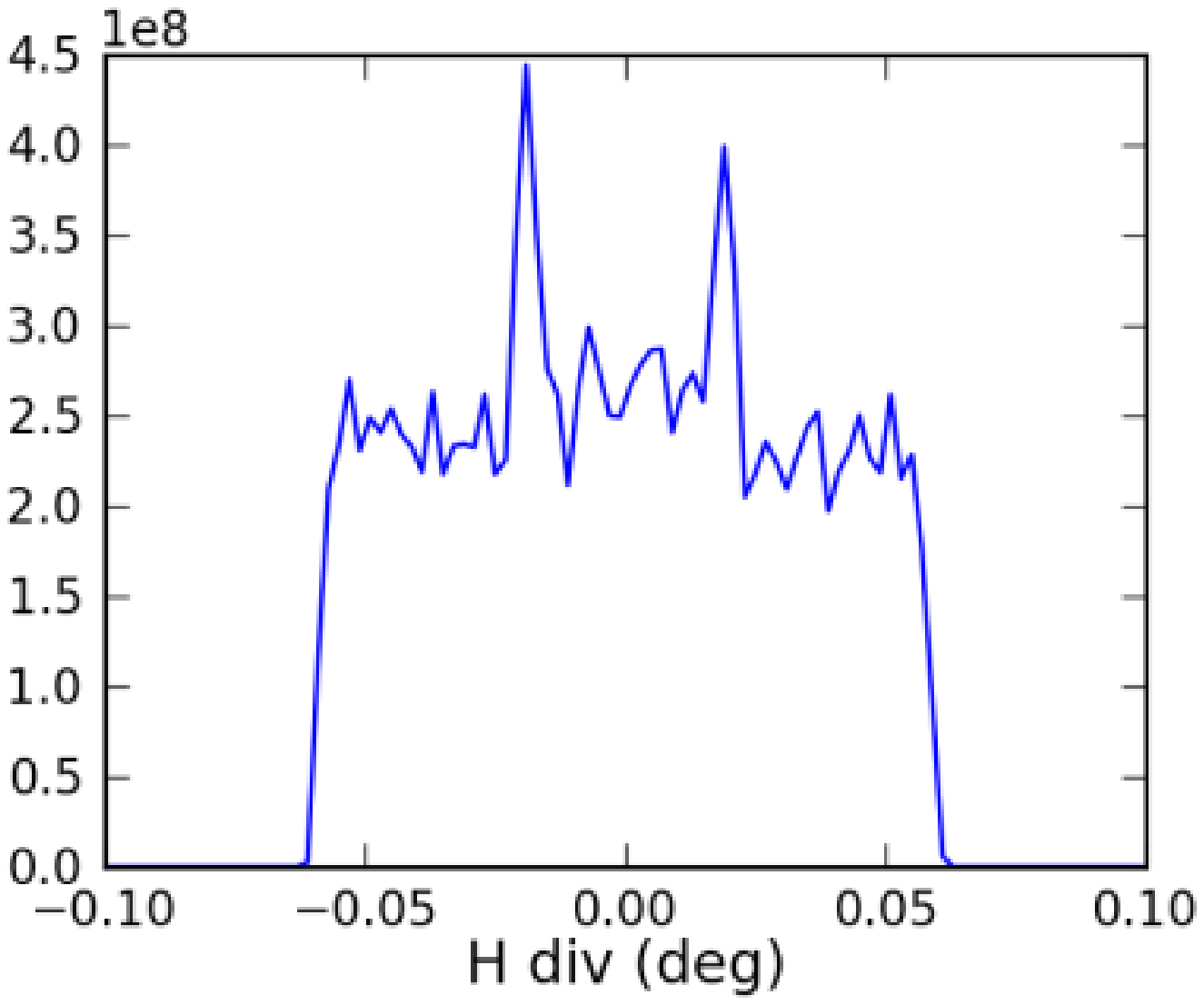}}
        \subfigure[]{\includegraphics[width=0.31\textwidth]{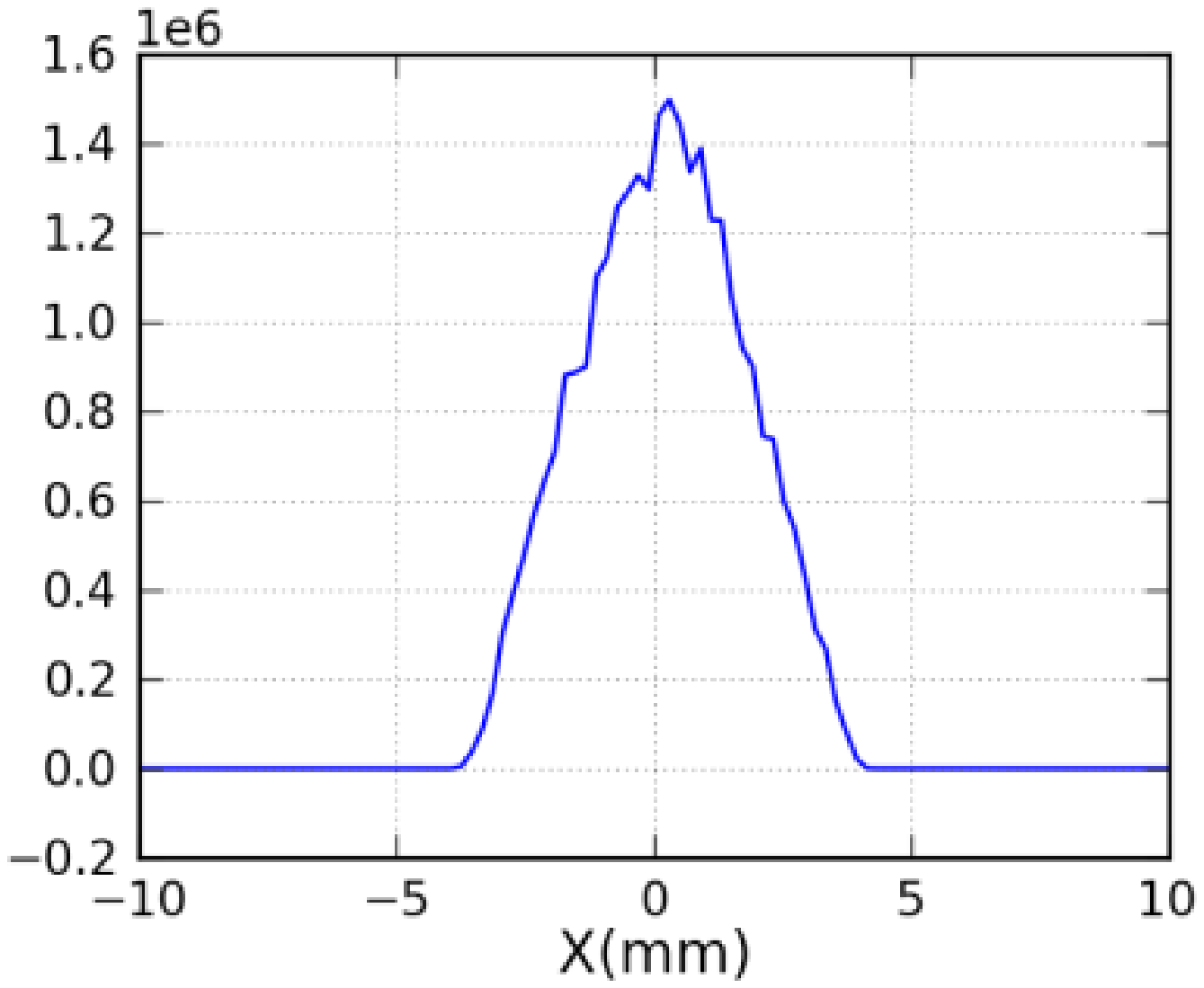}}
	\caption{Illustration of expected (a) intensity distribution in the horizontal plane at collimator exit, (b) horizontal divergence (full beam) at sample position, (c) profile of the beam at the detector position. The intensity is in arbitrary units.}
\label{Fig:GISANS1D}
\end{center}
\end{figure}

The alternative module for studies of laterally structured interfacial systems is a GISANS add-on. For this reason the reflectometer can also be equipped with a conventional GISANS module. The collimation concept of this module is based on the approach successfully used at REFSANS \cite{Bib:Kampmann}. In this concept the sample is illuminated by multiple low divergence beams which all converge to a single focus point on the detector (Fig. \ref{Fig:GISANSLayout}). In this way the best possible use of the available source divergence is realized and a high reciprocal space resolution is maintained. The radial collimator is replacing guide part 5 (bending section) and the collimation slits in front of the sample (see Fig. \ref{Fig:LayoutWFM} (b) and Tab. \ref{Tab:Reflectometer}). It consists of elements with non-reflecting side walls. Top and bottom walls are made from m=5 mirrors in order to provide sufficient beam bending for GISANS experiments. The collimator is divided laterally in three sub-channels separated by 0.1 mm thick absorbing walls. The chosen length L of the collimator is 6.4m with $w_1$ = 5 mm and $w_2$ = 3.3 mm. The detector is placed at 5.7 m downstream from the collimator exit with a sample-to-detector distance, LSD, of 5 m, slightly reducing the usable waveband by $\approx 5\%$. In the chosen 3 beams configuration, the intensity distribution is homogeneous horizontally (see Fig. \ref{Fig:GISANS2D}). The shadows of the collimator walls are small in comparison to the individual beam width. This ensures a very good sample illumination ($ > 90\%$). From Fig. \ref{Fig:GISANS1D} a) it appears that all three channels deliver the same intensity (resp. 30\%, 38\%, 32\% of the total), which ensures a homogeneous sampling of the surface. \\

The long collimation distance produces an intensity distribution that is uniform over the total horizontal divergence (Fig. \ref{Fig:GISANS1D}, left panel), and does not exceed +/-1mrad. As expected, each of the beams contributes equally to this divergence (Fig. \ref{Fig:GISANS1D}, middle panel). At the detector position, which was chosen to be 5 m from the sample in order not to depart too much from the normal operation distance of the reflectometer, one observes -as desired- a single spot of 4 mm full width at half maximum, matching the targeted in-plane resolution (Fig. \ref{Fig:GISANS1D}, right panel).
In conclusion the radial collimation option offers the possibility to perform GISANS measurements at the liquid reflectometer with a resolution matching that of a current state-of-the-art SANS instrument.

\section{Performance}

\begin{figure}
\begin{center}
        \includegraphics[width=0.98\textwidth,keepaspectratio]{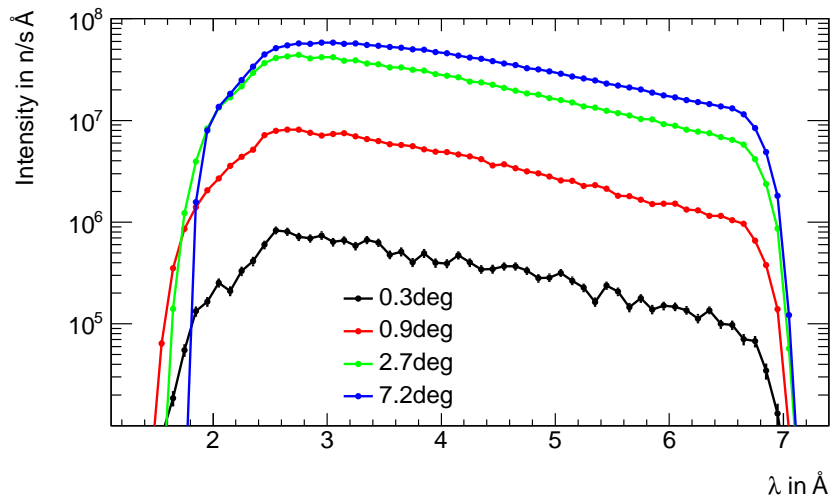}        
\caption{Beam intensity at the sample position as a function of wavelength $\lambda$ for several angle of incidence $\theta$. The neutron flux depends both on the diverence range, which is $\leq 10\%$ of the incident angle and the footprint size of the 1 cm$^2$ sample increasing with $\mathrm{sin} \, \theta$. See also Tab. \ref{Tab:Fluxes}.}
\label{Fig:FluxSample}
 \end{center}
\end{figure}

The layout of the horizontal reflectometer is optimized for measurements with small samples and high q transfers. Its performance is demonstrated by carrying out virtual experiments with different samples of 1x1 cm$^2$ area. Fluxes and count rates are given for different angular and resolution settings, which are summarized in Tab. \ref{Tab:Fluxes}. The beam intensity at the sample position is also shown for selected angular settings in Fig. \ref{Fig:FluxSample}. The simulations were carried out including the ESS cold moderator characteristics as of May 2013.

\subsection{Basic setup}

As discussed in the previous section, the basic setup of the reflectometer utilizes three choppers that provide the desired waveband with the wavelength resolution mainly determined by the length of the instrument going up to 10\% for 2 $\mathrm{\AA}$ neutrons. The slit collimation yields a vertical beam divergence that matches the loosest wavelength resolution (but is restricted to max. $\approx 0.6^{\circ}$ FHWM). Such a setup is most suitable for measuring reflectivity spectra of free liquid surfaces and monolayers on liquid surfaces, like e.g. Langmuir and Gibbs adsorption layer of amphiphiles. The measurement can be performed using the single waveband and three angular settings or the double waveband and two angular settings (and so on). In general, for each angular setting the waveband can be freely selected, e.g. for small angles it is advantageous to use a larger waveband, since the reflectivity is high at low q and thus a larger q-region can be covered with a measurement time still being of the order of a few seconds. The applicability will depend on the time required by the choppers to change settings from single to double (triple,...) waveband regime. \\

\begin{figure}
\begin{center}
	\subfigure[Reflectivity of a D$_2$O surface measured with the single waveband]{\includegraphics[width=0.47\textwidth]{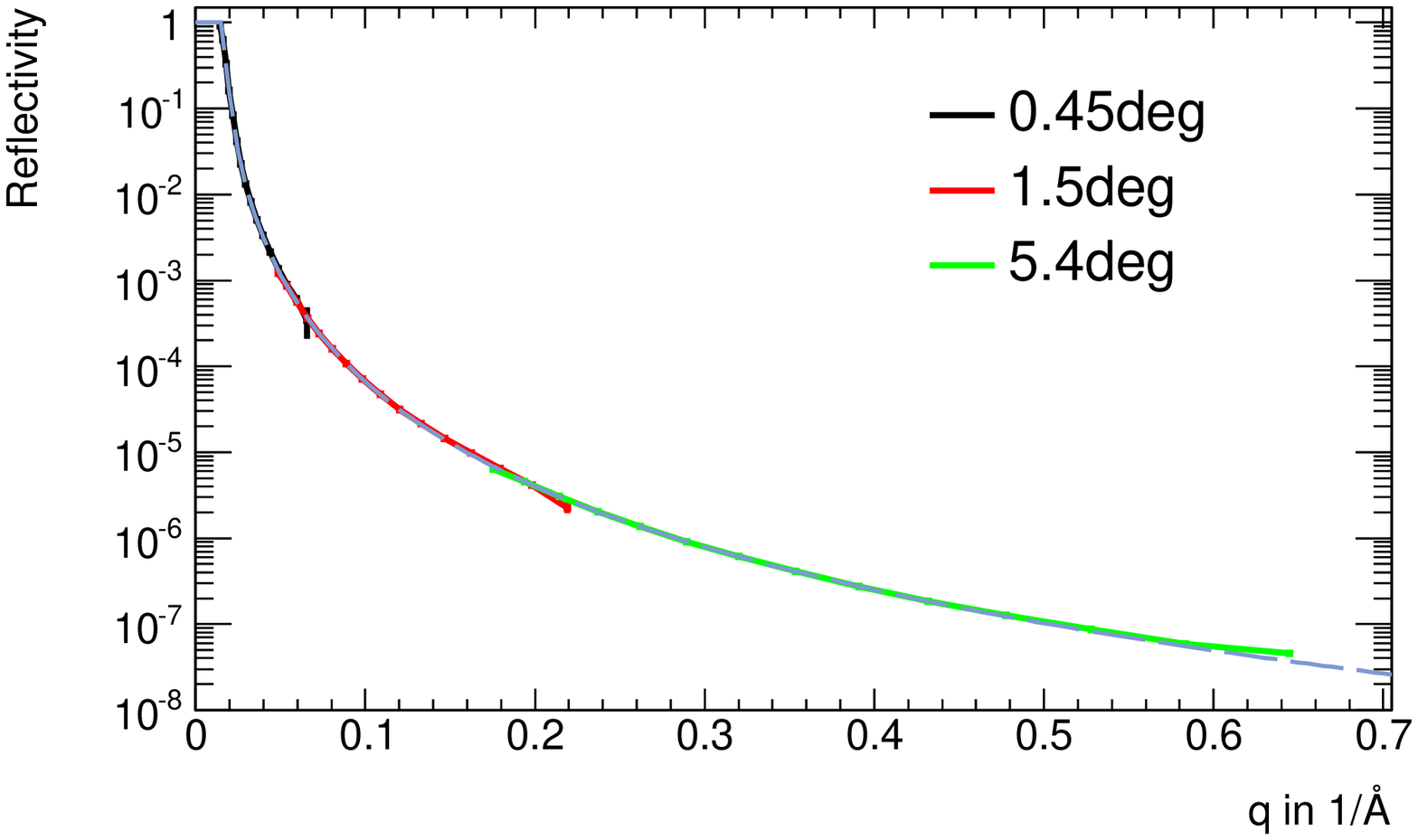}}
	\subfigure[Count rate achieved on a D$_2$O surface with the single waveband]{\includegraphics[width=0.47\textwidth]{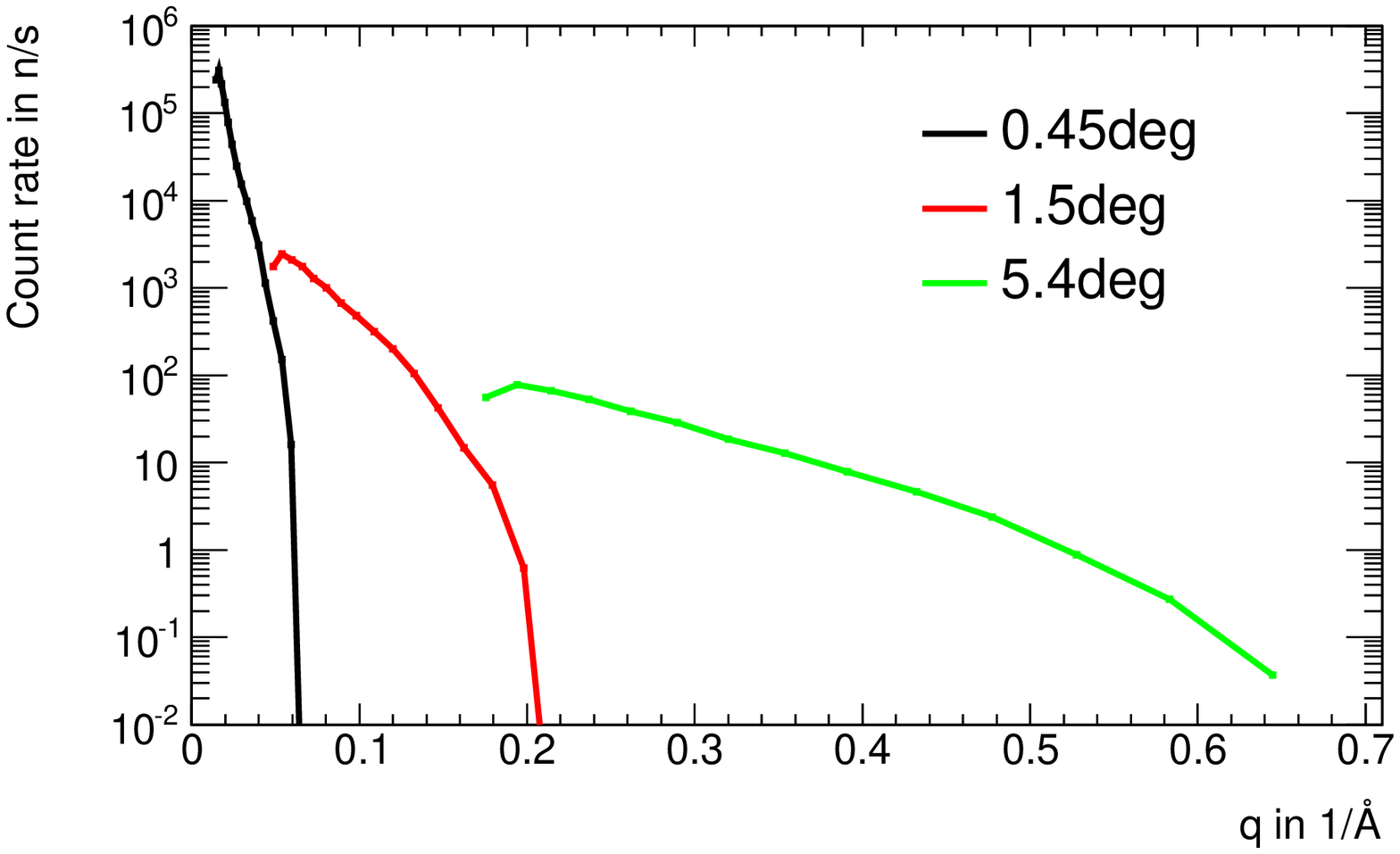}}
	\hfill
	\subfigure[Reflectivity of a D$_2$O surface measured with the double waveband]{\includegraphics[width=0.47\textwidth]{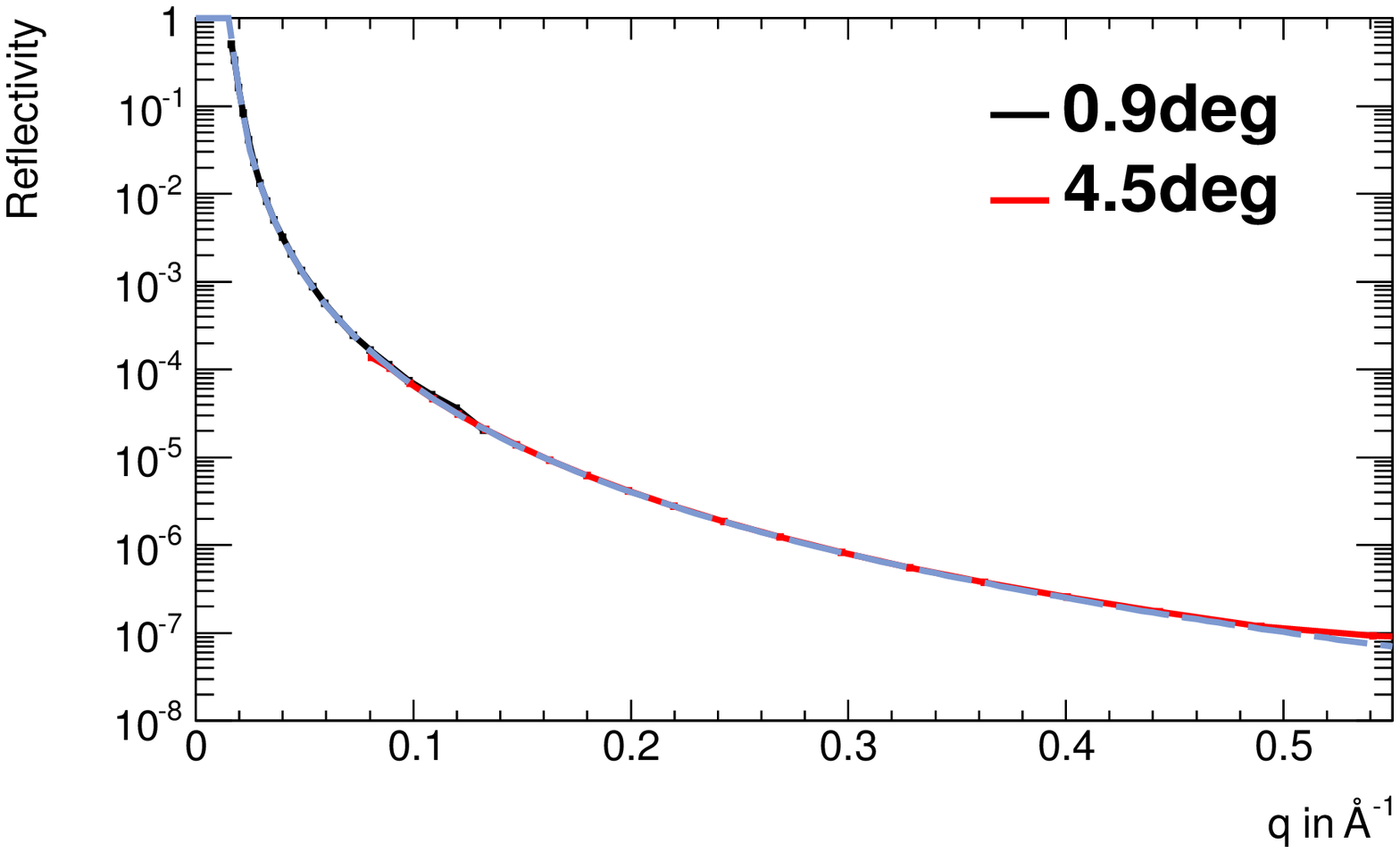}}
	\subfigure[Count rate achieved on a D$_2$O surface with the double waveband]{\includegraphics[width=0.47\textwidth]{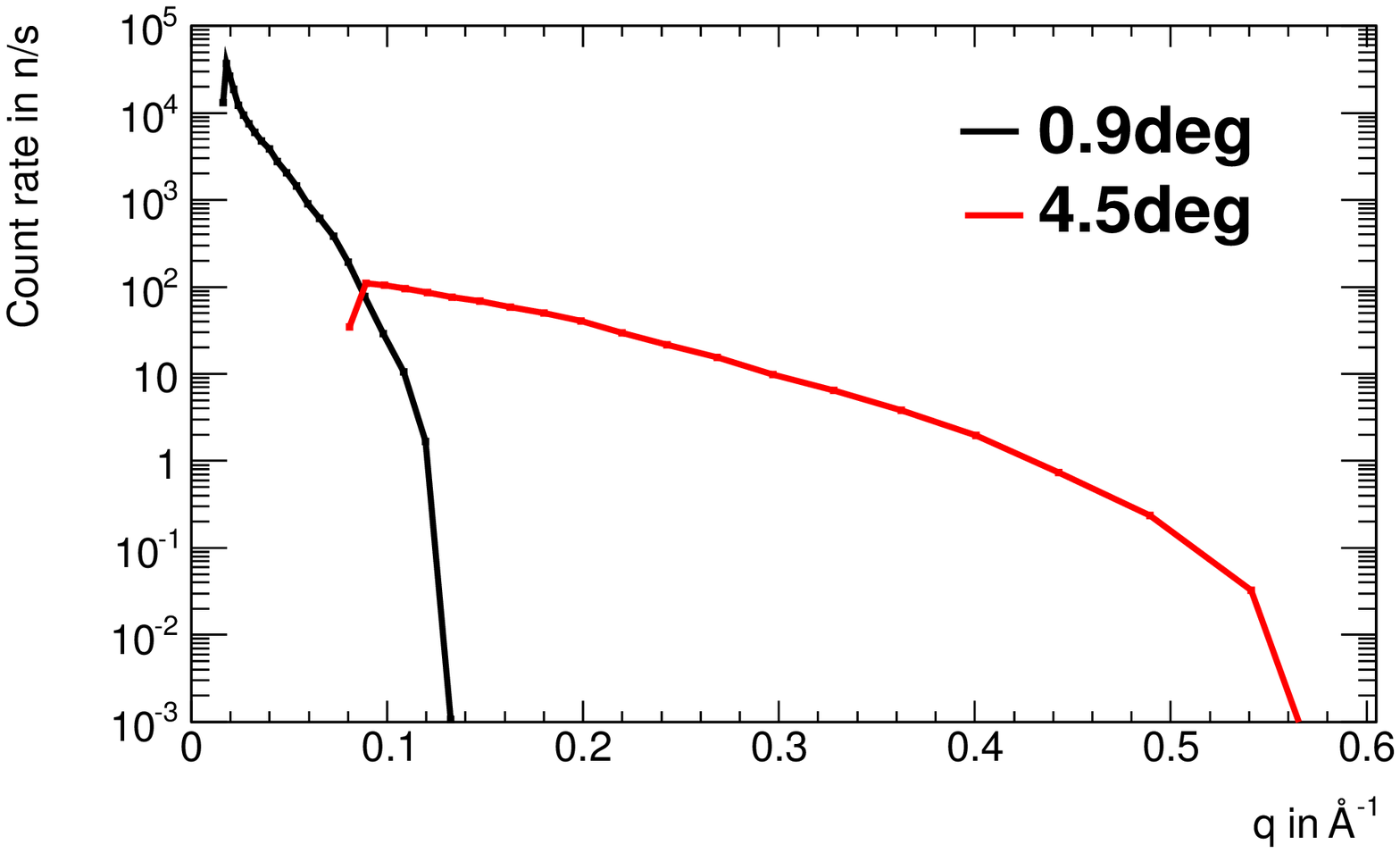}}	
	\hfill
	\subfigure[Reflectivity of a Langmuir layer on null reflecting water measured with the single waveband]{\includegraphics[width=0.47\textwidth]{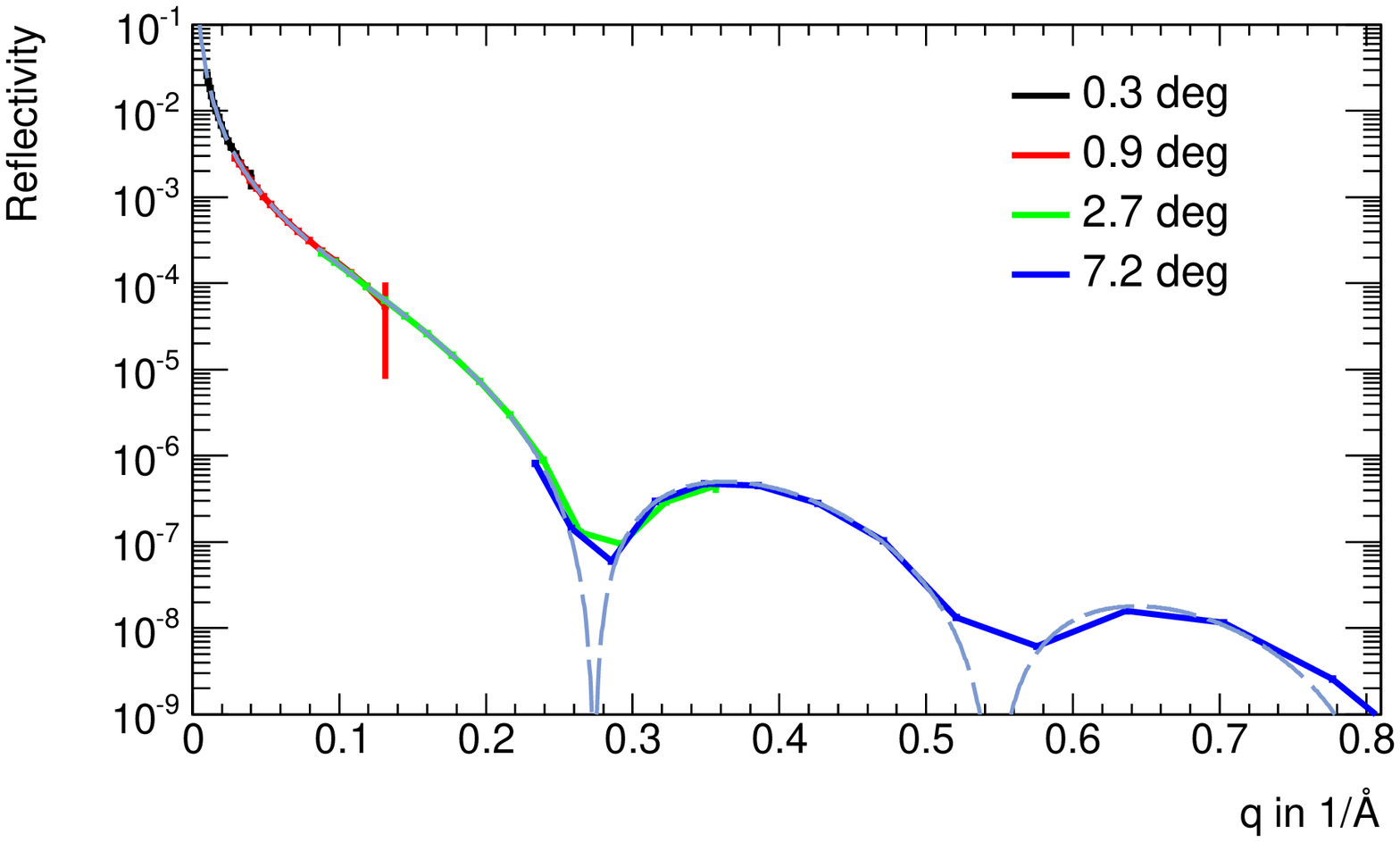}}
	\subfigure[Count rate achieved on a Langmuir layer surface with the single waveband]{\includegraphics[width=0.47\textwidth]{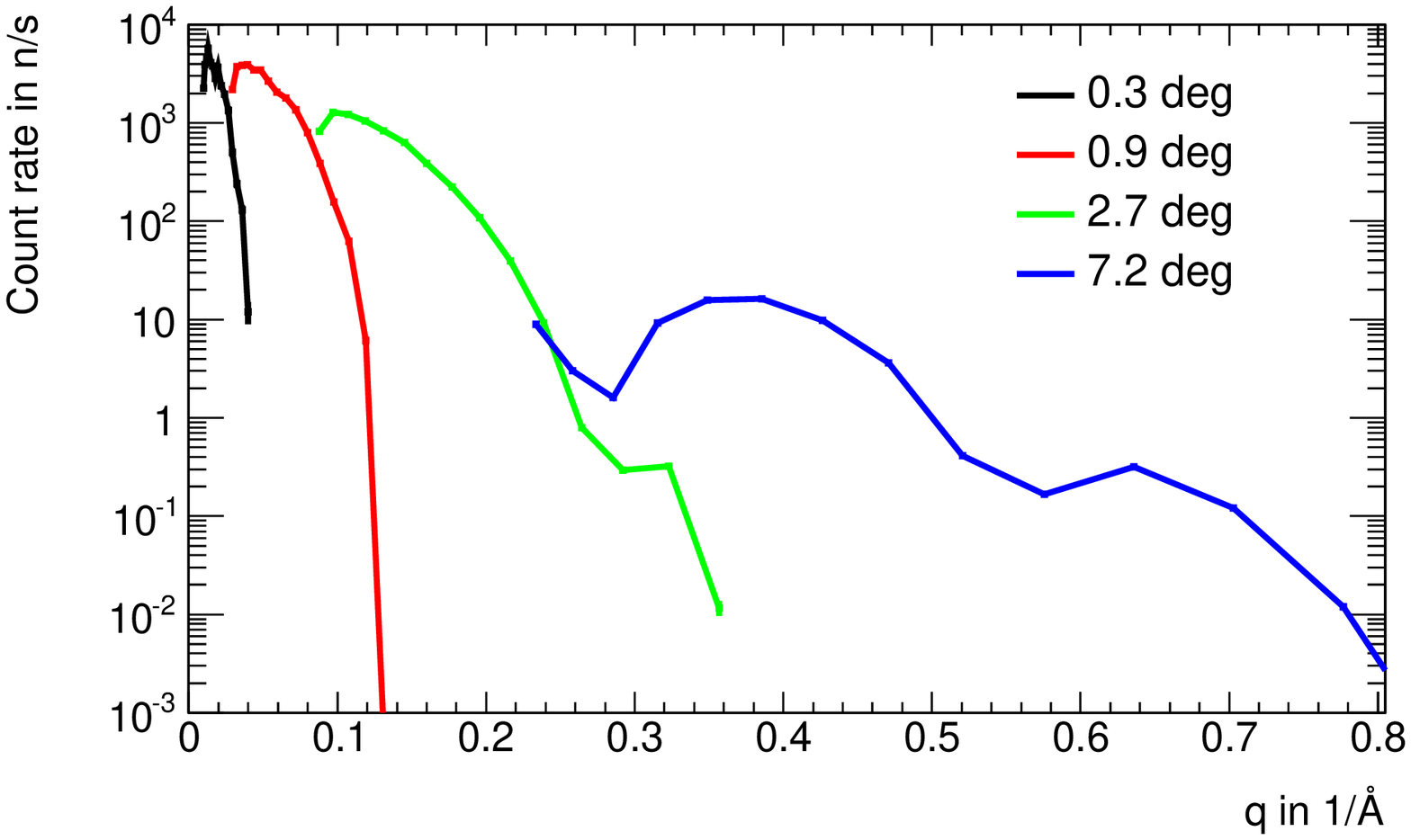}}
        \hfill
	\caption{a) - d): Virtual measurement of an ideal air-D$_2$O surface with the basic chopper setup and a $d \theta / \theta$ resolution of 10\%, using the single waveband and three angular settings and the double waveband and two angular setting, respectively. The count rate at the reflectivity of $10^{-7}$ is of the order of 1 n/s (0.25 n/s) for the single (double) waveband and thus 100 counts are collected within $\approx$ 1.5 (6) minutes for the largest angular setting.  The ideal reflectivity curve of the D$_2$O surface is shown in grey for comparison. e) - f): Virtual measurement of a Langmuir layer on null reflecting water reflectivity spectrum with the basic chopper setup and a $d \theta / \theta$ resolution of 10\%, using the single waveband and four angular settings. The count rate at the reflectivity of $10^{-8}$ for the highest reflection angle is of the order of 0.25 n/s, corresponding to a measurement time of 6 minutes for 100 signal counts. The ideal reflectivity curve of the d-DPPE Langmuir layer is shown in grey for comparison. Note that already in the basic setup the thickness of the Langmuir layer can be determined unambiguously, as two adjacent minima positions are resolved. }
\label{Fig:D20Measurement}
\end{center}
\end{figure}

The measured reflectivity curves of an ideal D$_2$O reflectivity surface and an adsorption monolayer\footnote{The free floating Langmuir layer is a condensed monolayer of 1,2-dipalmitoyl(d62)-sn-glycero-3-phosphoethanolamine (d-DPPE), chosen as a reference monolayer of amphiphilic molecules. The d-DPPE layer is 23 $\mathrm{\AA}$ thick, has an SLD of $5.63 \times 10^{-6} \, \mathrm{\AA}^{-2}$ and comes with rms roughness of 2 $\mathrm{\AA}$ at both the headgroup/liquid and the tailgroup/air interface. The parameters of the d-DPPE Langmuir layer are based on x-ray data \cite{Bib:Miller, Bib:Pabst, Bib:Wydro}.} and corresponding count rates are shown in Fig. \ref{Fig:D20Measurement}. The usage of the double waveband leads to a larger measurement time, but on the other hand the time needed to adjust for an intermediate reflection angle is saved. The longest measurement time is required, as expected, for highest angular settings. Nevertheless, it is possible to measure reflectivities up to $ q = 0.6 \, \mathrm{\AA}^{-1}$ within minutes on a 1x1 cm$^2$ sample, with at least 100 counts for every data point. The ability of measuring reflectivities at high q and on samples of 1 cm$^2$ size is unprecedented. For example, the FIGARO reflectometer at ILL is used for measurements up to 0.4 $\AA^{-1}$ with typically much larger sample sizes. \\

\begin{figure}
\begin{center}
	\includegraphics[width=0.98\textwidth]{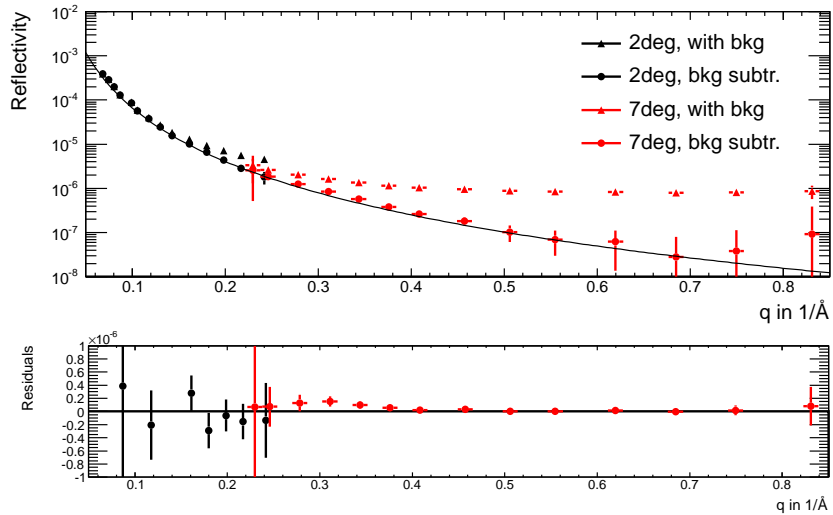}	
	\caption{The reflectivity curve of an ideal D$_2$O surface measured on the reflectometer including background from incoherent scattering. The original curve reaches its minimum at a reflectivity of $\approx 10^{-6}$. After subtracting the flat background (per angular setting) the measurement is in a good agreement with expectations. Note that the error bars correspond to the square root of the number of expected counts collected over 100 s.}
\label{Fig:InBkgMeasurement}
\end{center}
\end{figure}

It is well known that incoherent background originating from the sample sets a limit on minimum reflectivity that can still be measured with a certain statistical significance. For the ideal D$_2$O surface of 1x1 cm$^2$ area and 0.3 mm thickness, the background is of the order of $R_{\mathrm{BG}} = 10^{-6}$ using the mean free path of $\lambda_i = 7.353$ cm for D$_2$O for incoherent scattering. Neutronic simulations with incoherent background included show that even in that case data acquisition times of 100 s suffice in the range up to 0.6 $\mathrm{\AA}^{-1}$, see Fig. \ref{Fig:InBkgMeasurement}. Since its shape is flat, the background can be fitted and subtracted from the reflectivity spectrum. The residual spectrum agrees well with the theoretical curve and all data points still have a high enough statistical significance, with 2$\sigma$ being the significance of the lowest data point. Thus, if the measurement can be carried out for a sufficiently long time, being only a few minutes for a 1x1 cm$^2$ D$_2$O sample, reflectivities down to 10$^{-7}$ can be accessed by background recording and subtraction. In the case of the 1x1 cm$^2$ monolayer sample, the mean free path of the null reflecting water is  $\lambda_i = 0.202 \, \mathrm{cm}$ and yields a background level of $> 10^{-5}$. For such a sample the total counting time of the order of 4 to 5 hours  (for the largest angular setting) will suffice to achieve a statistically significant measurement of the monolayer signal after background subtraction. Naturally, the required sampling time is inversely proportional to the total sample area and can be significantly reduced if larger samples (of the order of 10 cm$^2$) are used.\\

\subsection{High-resolution setup}

\begin{figure}
\begin{center}
	\subfigure[The reflectivity of the NiTi sample]{\includegraphics[width=0.97\textwidth]{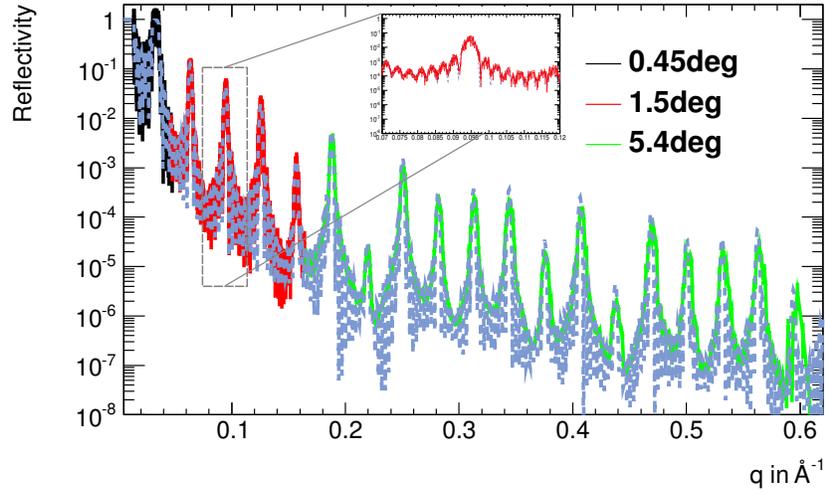}}
	\subfigure[The count rate for the measurement of the NiTi reflectivity]{\includegraphics[width=0.97\textwidth]{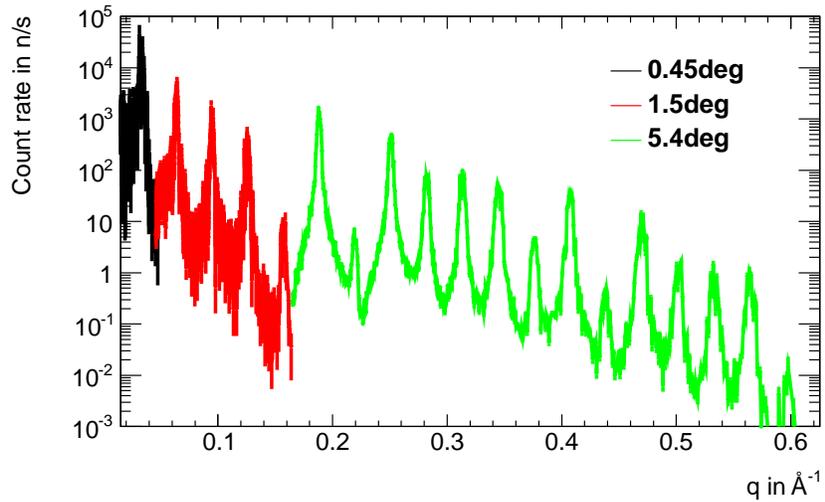}}
	\caption{The NiTi reflectivity curve measured on the reflectometer using the high-resolution WFM chopper setup. Even finer structures like Kiessig oscillations in the reflectivity spectrum can be well imaged. The measurement time for this sample would be around 4 min. For the last two angular settings, at least 25 counts per data point (taking into account the lowest count rate of 0.1 n/s) are collected in this way that is sufficient for a statistically significant measurement, if background can be neglected.}
\label{Fig:WFMMeasurement}
\end{center}
\end{figure}

The WFM chopper setup is optimized to provide a constant and high wavelength resolution for the single waveband from 2 $\mathrm{\AA}$ to 7.1 $\mathrm{\AA}$, combined with an adequate collimation before the sample for a high q-resolution. Due to a substantial loss in flux because of increased resolution as compared to the basic setup, see Fig. \ref{Fig:FluxBasicWFM}, the available q-range for such measurements is most likely reduced to $q < 0.5 \, \mathrm{\AA}^{-1}$. Within the accessible q-range, however, fast measurements of highly structured reflectivity spectra are rendered possible. In Fig. \ref{Fig:WFMMeasurement}, the measurement of a NiTi-multilayer sample ([86 $\mathrm{\AA}$ Ni/115 $\mathrm{\AA}$ Ti] on glass, total thickness = 2010 $\mathrm{\AA}$) is shown together with the ideal spectrum of this sample in grey for comparison. Its reflectivity spectrum exhibits several main peaks along with Kiessig oscillations over the entire q-range. The high-resolution setup of the reflectometer proves capable of a precise reconstruction of most of these features. To achieve at least 100 counts in each data point, 15 min of acquisition time for the highest angular setting is needed.

\begin{table}[htbp]
\begin{center}
  \begin{tabular}{|c|r|c|}
    \hline
    \textbf{Position $[\mathrm{m}]$} & \textbf{Description} & \textbf{Neutron flux} $[\mathrm{n \, / \, s \, cm^2}]$ \\ 
     \hline
     \hline
     2 & Guide entry & $7.3 \times 10^{10}$\\
     6 & End of extraction system & $3.6 \times 10^{10}$\\
     \hline
     \hline
     \multicolumn{3}{|c|}{Basic setup ($\delta \lambda / \lambda = 3\% - 10\%$, $ \delta \theta / \theta = 10\%$)} \\
     \hline
     18.7 & Before kink &  $7.7 \times 10^9$ \\
     29.9 & After kink &  $6.2 \times 10^9$ \\
     31 & After all choppers & $5.7 \times 10^9$ \\
     44.1 & Before bending section & $6.3 \times 10^9$ \\
     \hline
      &  $ \theta = 0.3^\circ$ & $3.1 \times 10^8$ \\
      & $ \theta = 0.45^\circ$ & $4.8 \times 10^8$\\	
      &  $ \theta = 0.9^\circ$ & $8.3 \times 10^8$ \\	
     52.5 & At footprint slit, $ \theta = 1.8^\circ$ & $9.9 \times 10^8$\\	
     & $ \theta = 2.7^\circ$ & $1.9 \times 10^9$ \\
      & $ \theta = 5.4^\circ$ & $1.7 \times 10^9$\\	
     &  $ \theta = 7.2^\circ$ & $1.3 \times 10^9$ \\
     \hline
     \multicolumn{3}{|c|}{High-resolution setup ($\delta \lambda / \lambda = 2.2\%$, $ \delta \theta / \theta = 2.2\%$)} \\  
     \hline	  
     18.7 & Before kink &  $2.0 \times 10^9$\\
     29.9 & After kink & $1.6 \times 10^9$  \\
     30.4 & After all choppers & $1.6 \times 10^9$ \\
     44.1 & Before bending section & $1.7 \times 10^9$\\
      \hline
      & $\theta = 0.45^\circ$ & $3.0 \times 10^7$\\
     52.5 & At footprint slit, $\theta =1.5^\circ$ & $6.0 \times 10^7$\\	
      &  $\theta =5.4^\circ$ & $2.3 \times 10^8$\\
     \hline
      \multicolumn{3}{|c|}{High-resolution setup ($\delta \lambda / \lambda = 1\%$, $ \delta \theta / \theta = 1\%$)} \\  
     \hline	  
     18.7 & Before kink &  $1.2 \times 10^9$\\
     29.9 & After kink & $9.4 \times 10^8$  \\
     30.4 & After all choppers & $9.1 \times 10^8$ \\
     44.1 & Before bending section & $9.9 \times 10^8$\\
      \hline
      & $\theta = 0.45^\circ$ & $7.5 \times 10^6$\\
     52.5 & At footprint slit, $\theta =1.5^\circ$ & $1.5 \times 10^7$\\	
      &  $\theta =5.4^\circ$ & $6.2 \times 10^7$\\
     \hline
  \end{tabular}
\end{center}
\caption{Neutron flux at different positions along the instrument for the basic and high-resolution setup, respectively. All values were obtained for the single waveband between 2 $\mathrm{\AA}$ and 7.1 $\mathrm{\AA}$ by averaging the flux either over the total guide cross-section at the corresponding position or over a $2 \times 1$ cm$^2$ rectangle perpendicular to the beam axis for measurements at the footprint slit (Slit 2). The flux at the footprint slit position is influenced both by the number of neutron reflections within the bending section and the opening of the first collimation slit (Slit 1) defining the divergence at the sample position.}\label{Tab:Fluxes}	
\end{table}

\section{Concept robustness}

The concept of the horizontal reflectometer makes mainly use of established and well known instrument components. The guide system consists of straight sections of 0.5 m. Solely the z-kink piece might require a somewhat smaller segmentation. The required coating is m=5 for the top and bottom guide surface, while except for the feeding section, the z-kink and the bending section, m=3 coating can be used everywhere else. The avoidance of the line of sight by the chosen guide geometry is assumed to clear the detector area from background arising from the prompt pulse. Two guide sections were studied more carefully, being the feeding and the deflection sections. The impact of various solutions for the feeding section located in the central monolith was inspected with respect to the delivered flux on sample. It was observed that even if the first 2 m of the feeding section cannot be installed due to cooling problems or other technical constraints, the flux on a 1x1 cm$^2$ sample stays without significant changes. Solely for samples with widths larger than 2 cm a drop of intensity occurs. The conclusion is that the instrument does not heavily depend on the actual performance of the extraction system. \\

The required stability of the deflecting guide system was studied with respect to the flux delivered on the sample. It was found that the misalignment of the individual guide pieces can be as large as 0.01$^{\circ}$ without causing noticeable flux losses. The precision that is routinely reached today is 0.001$^{\circ}$. Thus we conclude that the technical demands for the deflecting section do not pose any risk. \\

The chopper system is rather complex, but on the other hand the sizes of the chopper discs and their speed are well within what is technically feasible today. The practical validity of the WFM concept itself was already demonstrated at the BNC reactor in Hungary \cite{Bib:WFMExp}. Presuming that an adequate detector matching the instrument requirements is provided, the setup for high-resolution measurements does not entail additional risks. \\

\section{Discussion and conclusions}

The design of the horizontal reflectometer presented in this work has been developed with respect to scientific questions that will be relevant at the time when the ESS facility comes into operation. The horizontal reflectometer proves capable of measurements of high q transfers on small horizontal samples possible within reasonable measurement time, while being very flexible in terms of covered q-range, sample size and beam direction. A dedicated WFM chopper setup for high-resolution measurements on very thick multi-structured samples is currently the first one to be proposed for a reflectometer instrument and its design presented in this work and in \cite{Bib:WFM} shows that the required $\delta q / q$ resolution can be achieved without pushing chopper layouts at or beyond current technical limits. If this setup will be realized at the ESS, areas of parameter space that are currently inaccessible to neutron reflectometry due to limited flux and/or precision are expected to open up for exploration and new insights in materials organization.

\clearpage{}
\section*{Acknowledgements}

We thank our colleagues at the ESS for fruitful discussions.\\
This work was funded by the German BMBF under ``Mitwirkung der Zentren der Helmholtz Gemeinschaft und der Technischen Universit\"at M\"unchen an der Design-Update Phase der ESS, F\"orderkennzeichen 05E10CB1.''

\end{document}